\documentclass[%
 reprint,
superscriptaddress,
twocolumn,
nofootinbib,
 amsmath,amssymb,
 aps,
]{revtex4-2}
\usepackage{graphicx}% Include figure files
\usepackage{dcolumn}% Align table columns on decimal point
\usepackage{bm}% bold math
\usepackage{xcolor}
\usepackage{float}
\usepackage{soul}
\newcommand{\EuGa}{Eu(Ga$_{1-x}$Al$_x$)$_4$}

\begin{document}

%\preprint{APS/123-QED}

\title{Real-space and reciprocal-space topology in the Eu(Ga$_{1-x}$Al$_x$)$_4$ square net system}
%\textcolor{red}{Go through author list with EM and confirm affiliations with authors.}

\author{Jaime M. Moya}
\affiliation{Applied Physics Graduate Program, Rice University, Houston, TX, 77005 USA}
\affiliation{Department of Physics and Astronomy, Rice University, Houston, TX, 77005 USA}

\author{Jianwei Huang}
\affiliation{Department of Physics and Astronomy, Rice University, Houston, TX, 77005 USA}

\author{Shiming Lei}
\affiliation{Department of Physics and Astronomy, Rice University, Houston, TX, 77005 USA}

\author{Kevin Allen}
\affiliation{Department of Physics and Astronomy, Rice University, Houston, TX, 77005 USA}

\author{Yuxiang Gao}
\affiliation{Department of Physics and Astronomy, Rice University, Houston, TX, 77005 USA}

\author{Yan Sun}
\affiliation{Shenyang National Laboratory for Materials Science,Institute of Metal Research, Chinese Academy of Sciences}

\author{Ming Yi}
\affiliation{Department of Physics and Astronomy, Rice University, Houston, TX, 77005 USA}

\author{E. Morosan}
\email{em11@rice.edu}
\affiliation{Department of Physics and Astronomy, Rice University, Houston, TX, 77005 USA}

\date{\today}

\begin{abstract}
Magnetotransport measurements on the centrosymetric square-net \EuGa~ compounds reveal evidence for both reciprocal- and real-space topology. For compositions $0.50 \leq x \leq 0.90$, several intermediate field phases are found by magnetization measurements when $H \parallel c$, where a maximum in the topological Hall effect (THE) is observed, pointing to the existence of  topological (real-space topology) or non-coplanar spin textures.  For $0.25 \leq x \leq 0.39$, magnetization measurements reveal an intermediate field state, but no transition is visible in the Hall measurements. For $x = 0.15$, only one magnetic transition occurs below the N\'eel temperature $T_N$, and no intermediate field spin reorientations are observed. The Hall effect varies smoothly before the spin-polarized (SP) state.  However, in the SP state, Hall measurements reveal a large anomalous Hall effect (AHE) for all compositions, a consequence of reciprocal-space topology.  Density functional theory calculations in the paramagnetic state indeed reveal a Dirac point that lies very near the Fermi energy, which is expected to split into Weyl nodes in the SP state, thereby explaining the observed AHE.  These results suggest the \EuGa~ family is a rare material platform where real- and reciprocal-space topology exist in a single material platform.

\end{abstract}

\maketitle

\section{Introduction}

The advent of topological materials has generated much interest not only from a fundamental physics perspective, but also due to their potential applications in revolutionary electronic devices. For example, the theoretical prediction of the quantum anomalous Hall effect (QAHE) \cite{haldane1988model,liu2016quantum}, a consequence of Berry curvature \cite{berry1984quantal} in reciprocal space, promises the possibility of quantized, chiral, dissipation-free electron transport without magnetic fields, ideal for energy efficient electronic devices.  The reality of such a device was brought one step closer with the experimental realization of the QAHE in the topological insulator Cr-doped Bi(Sb)$_2$Te$_3$ \cite{chang2013experimental}.

The notions of topology in condensed matter systems also extends to real space via topological spin textures which are particle-like, non-coplanar spin configurations characterized by a  topological charge \cite{tokura2020magnetic}. Topological spin textures have been proposed for applications in next-generation memory, logic, spintronic and neuromorphic computing devices \cite{fert2013skyrmions, fert2017magnetic, song2020skyrmion,jonietz2010spin,nagaosa2013topological}. Furthermore, itinerant electrons coupled to topological spin textures acquire Berry phase when traversing the non-coplanar spin textures. Therefore, the topological spin textures can be regarded as real-space sources of Berry curvature, resulting in topological the Hall effect (THE) \cite{tokura2020magnetic}.

Combining reciprocal- and real-space topology extends the phase space of approaches for future spintronic applications. For example, proof-of-principle experiments  based on topological insulator/magnetic heterostructures exhibit both THE and QAHE: this approach takes advantage of THE to read out the spin-state information, which, in turn, can be transmitted without dissipation using the QAHE chiral edge states \cite{jiang2017skyrmions,jiang2020concurrence,li2022interplay,li2020topological,zou2022enormous,xiao2021mapping}. Compared to their heterostructure counterparts, bulk magnetic topological materials offer the chance for stronger coupling between magnetism and itinerant electrons. However, the concurrence of reciprocal- and real-space topology has not yet been observed in a bulk system.

The \EuGa~ series can be considered an ideal platform for such reciprocal- and real-space topology coexistence. The $x$ = 0 compound EuGa$_4$ orders magnetically into a simple A-type antiferromagnetic (AFM) structure \cite{kawasaki2016magnetic}. No intermediate phases are observed below the N\'eel temperature in the magnetic field - temperature ($H~-~T$) phase diagram \cite{Kevin2021}. However, above $T_N$, angle-resolved photoemission spectroscopy (ARPES) data show the existence of four-fold degenerate spinless nodal rings (NR) near the Fermi-level in EuGa$_4$ \cite{Kevin2021}. With the application of a magnetic field $H~\parallel~c$, when the magnetization is saturated in the spin-polarized (SP) state, the spin-degeneracy is lifted, and two topological Weyl-nodal rings (NRs) are realized, protected by mirror symmetry  \cite{Kevin2021}. The topological NRs, whose signatures are observed in magneto-transport measurements \cite{Kevin2021}, are responsible for the large quantum mobility and large, unsaturated magnetoresistance (MR) reaching a value of 200,000\% at $\mu_0H~=~14~$T and $T~=~2~$K \cite{Kevin2021,zhang2021giant}. The $x$ = 1 compound EuAl$_4$ \cite{stavinoha2018charge, takagi2022square,shang2021anomalous,meier2022thermodynamic,kaneko2021charge,shimomura2019lattice} has a complex $H - T$ phase diagram with several spin reorientation transitions for $H \parallel c$ \cite{meier2022thermodynamic,shang2021anomalous,takagi2022square}. A non-zero THE in select regions of the $H~-~T$ phase diagram led to the proposal of topological spin-textures in EuAl$_4$ \cite{shang2021anomalous}. Later, the existence of a skyrmion lattice was confirmed via small-angle nuetron scattering measurements \cite{takagi2022square}. The evidence for the reciprocal-space topology in EuGa$_4$ and  real-space topology in EuAl$_4$ prompts the search for the coexistence of the two types of topological states across the \EuGa~series.

Structurally, \EuGa crystallizes in the tetragonal space group $I4/mmm$, which hosts two crystollographic sites for Al or Ga \cite{stavinoha2018charge}.  For $x$ = 0.5, the Al and Ga preferentially occupy the two different sites, leading to an ordered structure of EuGa$_2$Al$_2$ \cite{stavinoha2018charge}. EuGa$_2$Al$_2$ has also been shown to have a complex $H - T$ phase diagram for $H \parallel c$, with a non-zero THE maximum centered around an intermediate-field phase, pointing to the existence of either a topological spin texture or another non-coplanar spin texture \cite{moya2022incommensurate}. The exact nature of this intermediate field state is yet to be determined.

The existence of reciprocal- and real-space topology for the two end members EuGa$_4$ and EuAl$_4$, respectively, and the persistence of the THE in EuGa$_2$Al$_2$, motivate a  detailed study on the \EuGa~ series, with a goal of identifying compositions where both topological phenomena might coexist. To this end, \EuGa~ single crystals with $x$ = 0.15, 0.24, 0.31, 0.39, 0.50, 0.58, 0.71 and 0.90 were synthesized.  In the magnetically ordered state, the THE for $x \geq 0.50$ with $H \parallel c$ is observed, pointing to the existence of topological or, more generally, non-coplanar spin textures. Additionally, a large intrinsic AHE is registered in the field-induced spin-polarized (SP) state when the magnetization is saturated for compositions 0.24 $\leq$ x $\leq$ 0.71, suggesting reciprocal-space band topology. Density functional theory (DFT) calculations indicate the existence of a Dirac point near the Fermi energy in EuGa$_2$Al$_2$.  The Dirac point is expected to split into Weyl nodes in the SP state, responsible for the large AHE in \EuGa.

\section{Experimental Methods}

Single crystals of \EuGa~ were synthesized using a self-flux method described in Ref.~\cite{stavinoha2018charge}. Powder X-ray diffraction measurements were collected with a Bruker D8 Advance diffractometer with Cu K$_\alpha$ radiation. Rietveld refinements were done using Fullprof software \cite{rodriguez1993recent} and the obtained lattice parameters are consistent with the previous results \cite{stavinoha2018charge}.  
Quantitative elemental analysis by Wavelength Dispersive Spectrometry (WDS) of \EuGa~ phase was performed using the EPMA (Electron Probe Micro-Analyzer) instrument at Rice University, with a JEOL JXA 8530F Hyperprobe equipped with a field emission (Schottky) emitter and five WDS spectrometers. The analytical conditions used were 15 kV accelerating voltage, 20 nA beam current and a spot beam size of $\sim$ 300 nm. The standards used for composition calibration were synthetic in-house produced stoichiometric compounds EuGa$_4$ and EuAl$_4$. Careful background offsets were manually selected for each element to avoid interferences with higher order X-rays during peak and background measurement. Each element (Ga, Al and Eu) was simultaneously analyzed on two spectrometers, in order to improve the statistics on the standard deviation and detection limit calculation for each measurement. The reproducibility of the standards was accurate and precise, with an error below 1\% for each element.  ZAF matrix correction was employed for quantification.

% The X-ray analyzed lines, the analyzing crystal used, peak positions, background offsets, analytical standard deviation and detection limit for each element are shown in Table \ref{tab1} in the Supplementary Materials. The counting times of X-rays, and detector conditions are presented in Table \ref{tab2}. ZAF matrix correction was employed for quantification.

% The composition of EuGa$_2$Al$_2$ ($x$ = 0.5) was also determined using energy-dispersive X-ray spectroscopy (EDS) measurements, with consistent results between EPMA and EDS. EDS measurements were performed in a FEI Helios NanoLab 660 Dual Beam system with incoming electron energy 15 keV and intensity 0.2 nA. The spectra were analyzed in the software AZtecLive to obtain the atomic percentage of each element. All other compositions besides $x$~=~0.5 were subsequently determined by EDS.

Magnetization measurements were done using a Quantum Design (QD) Dynacool system equipped with a vibrating sample magnetometer. Four-probe resistivity measurements were made using the electrical transport option in the same system  with the typical applied current $j~=~5~$mA and frequency $f~=~9.15~$Hz. Measurements of the longitudinal resistivity as a function of magnetic field $\mu_0H$, $\rho_{xx}(\mu_0H)$, and Hall measurements, $\rho_{yx}(H)$, were measured in a complete field-sweep loop with four quadrants: Quadrant I from $\mu_0H~>~0$ to $\mu_0H~=~0$, Quadrant II from $\mu_0H~=~0$ to $\mu_0H~<~0$, Quadrant III from $\mu_0H~<~0$ to $\mu_0H~=~0$ and Quadrant IV from $\mu_0H~=~0$ to $\mu_0H~>~0$. The subsequent measurements were symmetrized or antisymmetrized, respectively. No hysteresis was observed for any composition.

% \textcolor{red}{Yan Please check this.}
% \textcolor{red}{DFT calculations were performed using the code of Vienna ab-initio simulation package (VASP) \cite{kresse1996efficient}. Experimental lattice parameters and atomic positions were used as the input.  The localized f-electrons, were accounted for with an on-site Hubbard  $\text{U}=5\,$eV which  was applied on Eu-4f orbitals \cite{dudarev1998electron}. The calculated magnetic moment is $~\sim$ 6.9 $\mu_\text{B}$/Eu, close to the expected value for Eu$^{2+}$. To check the DFT electronic band structure and magnetization, the full-potential local-orbital code (FPLO) with localized atomic basis and full potential \cite{koepernik1999full}, were used. }

The electronic band structure was calculated based on density functional theory by using the code of Vienna Ab initio Simulation Package (VASP) \cite{kresse1996efficient} with projected augmented wave potential. The exchange and correlation energies were considered at the level of the generalized gradient approximation (GGA), following the Perdew–Burke–Ernzerhof parametrization scheme \cite{perdew1996generalized}. The energy cut-off was set as 500 eV. The calculations have dealt with f-electrons as the valance states. To calculate the surface state, we projected the Bloch wavefunctions into maximally localized Wannier functions (MLWFs) \cite{mostofi2008wannier90} derived from the Eu-5d, Eu-6s, Eu-6p, Ga-4s, Ga-4p, Al-3s, and Al-3p orbitals. The tight-binding model Hamiltonian was constructed from the MLWF overlap matrix. Based on the tight-binding Hamiltonian, the surface state was considered under open boundary conditions with the half-infinite two-dimensional model using the iterative Green’s-function method \cite{sancho1984quick,sancho1985highly}.

\begin{figure*}
\includegraphics[width=\textwidth]{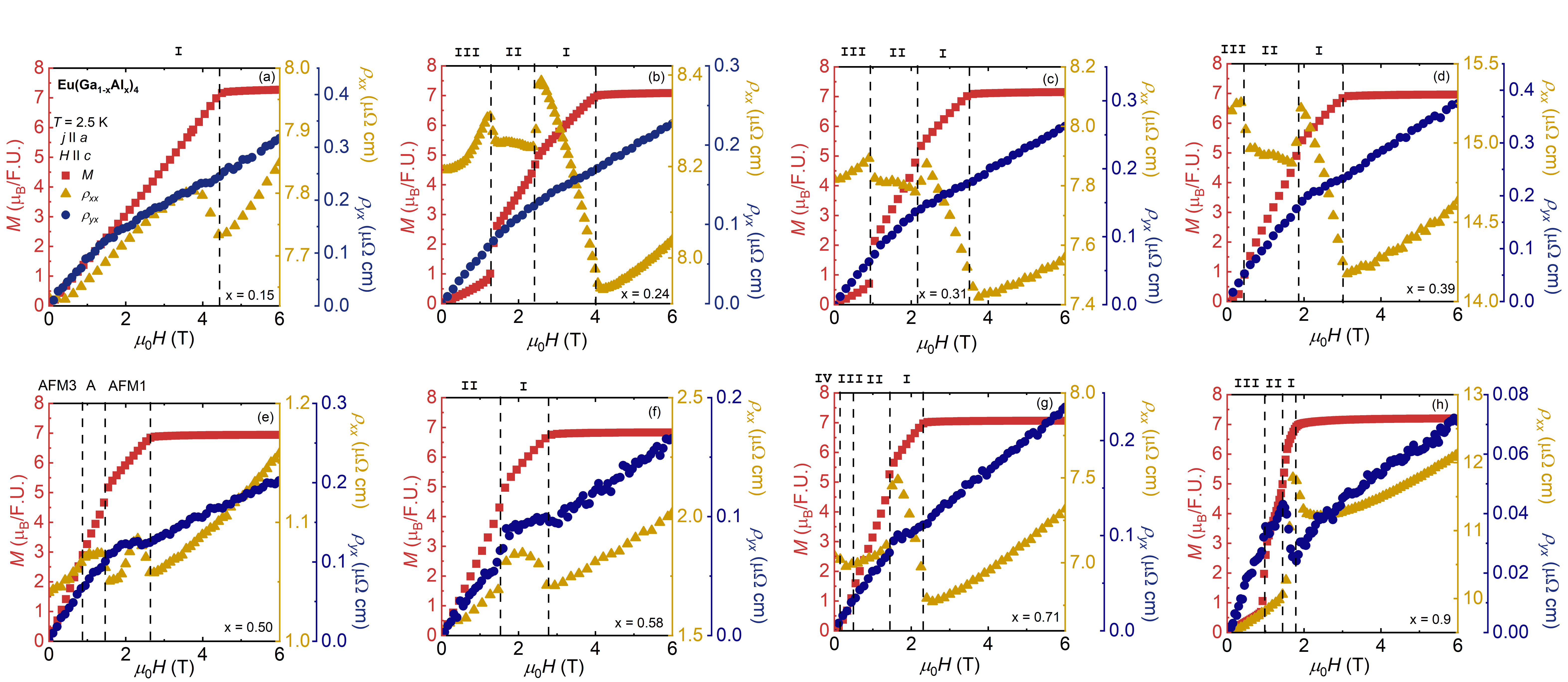}
\caption{\label{two5} Magnetization $M$ (red squares, left axis), resistivity $\rho_{xx}$ (gold triangles, inner right axis), and Hall resistivity $\rho_{yx}$ (blue circles, outer right axis) as a function of magnetic field $\mu_0H$  measured with magnetic field $H~\parallel~c$ at temperature $T~=~2.5~$K for \EuGa~(a) $x$ = 0.15, (b) $x$ = 0.24, (c) $x$ = 0.31, (d) $x$ = 0.39, (e) $x$ = 0.50, (f) $x$ = 0.58, (g) $x$ = 0.71 and (h) $x$ = 0.90. The transport measurements are measured with the current $j~\parallel~a$.  Phase boundaries are marked with a dashed line.}
\end{figure*}

\section{Results and Discussion}

\label{RandD}

Hall effect measurements have proven a powerful tool to identify systems with large sources of Berry curvature $\mathbf{\Omega_k}$ \cite{ye2018massive, suzuki2016large, liu2018giant,yang2020giant, nayak2016large,kurumaji2019skyrmion,fujishiro2021giant,lee2007hidden, liang2018experimental, singha2019magnetotransport, chen2021large,pavlosiuk2020anomalous,zhu2020exceptionally,lee2009unusual,shang2021anomalous,nagaosa2010anomalous,tokura2020magnetic,xiao2010berry}. Systems with non-zero $\mathbf{\Omega_k}$ have an additional term in the Hall conductivity $\sigma_{xy}$ besides the normal metal component, and this is the anomalous Hall conductivity $\sigma_{xy}^A$ that depends on $\mathbf{\Omega_k}$ as $\sigma_{xy}^A~=~-(e^2/h)\int d^3k\Omega_k^z/(2\pi)^3$ \cite{nagaosa2010anomalous}.  Berry curvature can either be generated via the Karpus-Luttinger-type (KL) mechanism \cite{karplus1954hall}, first discussed in the context of ferromagnets, or the scalar spin chirality (SSC) \cite{ohgushi2000spin,shindou2001orbital,martin2008itinerant} mechanism. In the former, spin-orbit coupling (SOC) has always been emphasized as a key ingredient while the derivation of the quantum Hall conductivity by Thouless \textit{et al}. \cite{thouless1982quantized} and subsequent interpretation of the non-quantized anomalous Hall conductivity by Haldane \cite{haldane2004berry} made clear the relationship between reciprocal-space topology and the anomalous Hall effect. In the latter, when the SSC defined as $\chi_{ijk}~=~\mathbf{S}_i\cdot(\mathbf{S}_j\times \mathbf{S}_k)$ (where $\mathbf{S}_{i,j,k}$ are the spins of three adjacent sites in a lattice) is non-zero, the real-space non-coplanar spin textures act as sources of Berry curvature. In the literature the Hall response contributed from the SSC-type mechanism is often referred to as the topological Hall effect. Since topological spin textures are non-coplanar, they should therefore exhibit a topological Hall effect if they are metallic \cite{tokura2020magnetic,kurumaji2019skyrmion,hirschberger2019skyrmion}.

\subsection{Topological Hall Effect}
\label{SS1}

Since both types of Hall effects are related to the magnetic properties, field-dependent magnetization data $M(H)$ for \EuGa~ (blue squares, Fig.~\ref{two5}) are compared with resistivity measurements $\rho_{xx}$ (gold triangles) and Hall resistivity $\rho_{yx}$ (blue circles), measured at $T = 2.5~K~<~T_N$, with $H \parallel c$, and current $j \parallel a$. Signatures of field-induced spin reorientation are clearly observed by anomalies in both $M$ and $\rho_{xx}$, which are marked by the vertical dashed lines in Fig.~\ref{two5}. Qualitatively, the behavior of $\rho_{yx}$ below the fields $\mu_0H_c$ where $M$ saturates can be grouped into two categories depending on composition, as discussed below.

\begin{figure*}
\includegraphics[scale = 0.3]{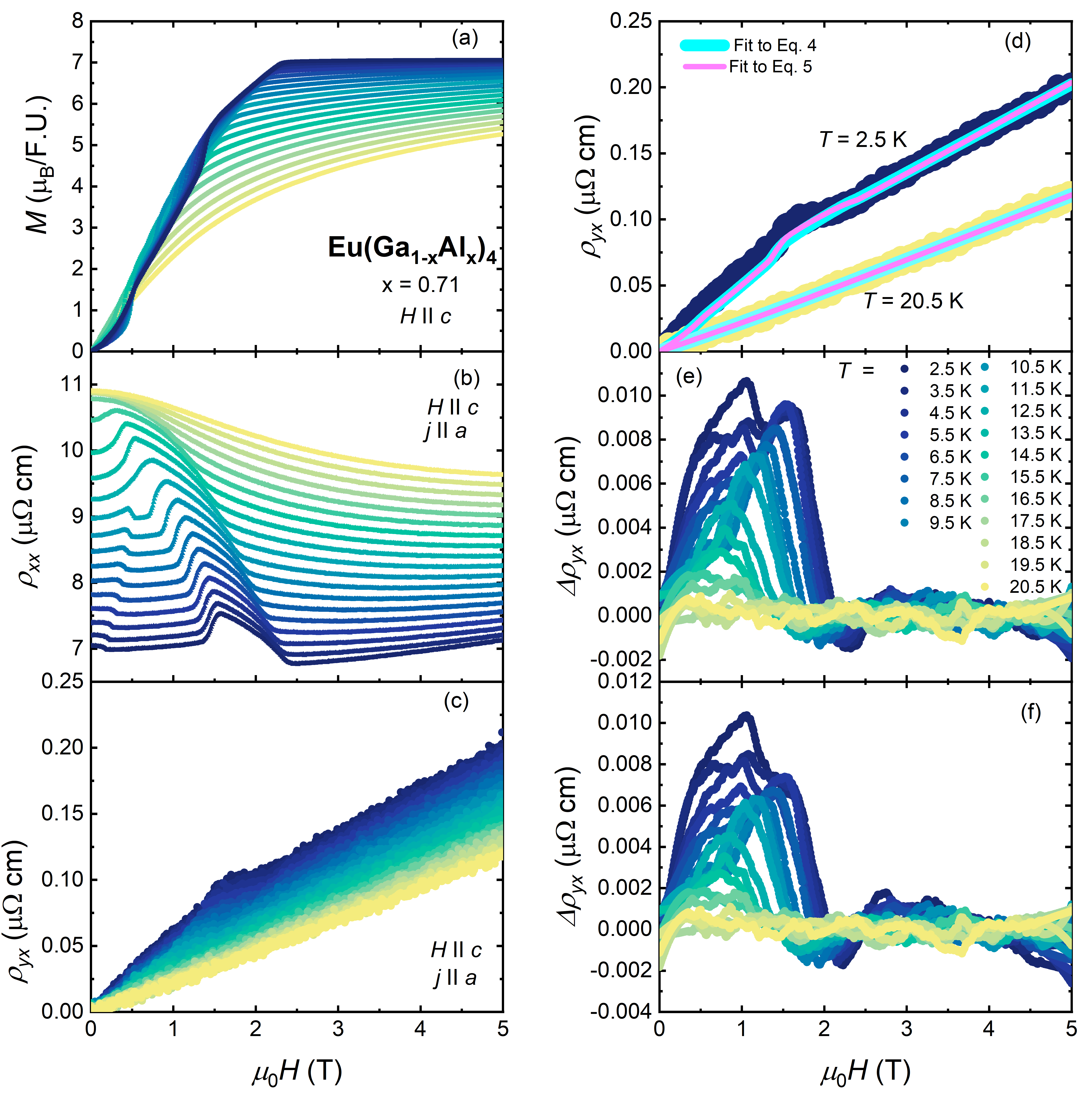}
\caption{\label{seventyone} Topological Hall effect for \EuGa with $x=~0.71$. (a) Isothermal magnetization $M$, (b) longitudinal resistivity $\rho_{xx}$, and (c) Hall resistivity $\rho_{yx}$ measured as a function of magnetic field $\mu_0H$ with field $H~\parallel~c$ at temperatures $2.5~\text{K}~\leq~T~\leq~20.5~$K.  Transport measurements are measured with current $j~\parallel~a$. (d) $\rho_{yx}$ at $T~=~2.5~$K (blue circles) and $T~=~20.5~$K (yellow circles) with fits to Eq.~\ref{Eq:skewII} (cyan lines) and Eq.~\ref{Eq:intrinsicII} (pink lines). (e) and (f) The difference, $\Delta\rho_{yx}$, between $\rho_{yx}$ and fits to Eq.~\ref{Eq:skewII} and Eq.~\ref{Eq:intrinsicII}, respectively. }
\end{figure*}

For $0.15 \leq x \leq 0.39$ (Fig.~\ref{two5}a-d), $\rho_{yx}$ is concave down below $\mu_0H_c$ with no obvious discontinuities even though there are clear magnetic transition features in both $M$ and $\rho_{xx}$ curves. For $x=0.15$ (Fig.~\ref{two5}a), only one magnetic phase transition is observed, which is similar to the behavior of the end-compound EuGa$_4$, where the field induces a magnetic transition from the AFM ground state to the fully SP state \cite{nakamura2013fermi}. It is therefore unlikely for this composition to host topological spin textures. For $0.24 \leq x \leq 0.39$ (Fig.~\ref{two5}b-d), the magnetic field drives the system through multiple magnetic phases, which are clearly captured in $M$ and $\rho_{xx}$, but not very evident in the $\rho_{yx}$ curves. 

For $0.50 \leq x \leq 0.90$ (Fig.~\ref{two5}e-h), $\rho_{yx}$ shows discontinuities corresponding to some of the magnetic transitions that are also revealed in $M$ and $\rho_{xx}$. The discontinuity in $\rho_{yx}$ suggests the existence of the THE, which is observed in many well established skyrmion hosting materials, such as MnSi \cite{neubauer2009topological,lee2009unusual}, Gd$_2$PdSi$_3$ \cite{kurumaji2019skyrmion}, Gd$_3$Ru$_4$Al$_12$ \cite{hirschberger2019skyrmion}, and GdRu$_2$Si$_2$ \cite{khanh2020nanometric}. Such a discontinuity is associated with the first-order nature of the field-induced magnetic transition. Therefore it is concluded that the \EuGa~ compounds with 0.5 $\leq~x~\leq$ 0.90 are promising candidates to host real-space topological spin textures. Next the THE for these compositions is  evaluated.

\begin{figure}[h!]
\includegraphics[width=\columnwidth]{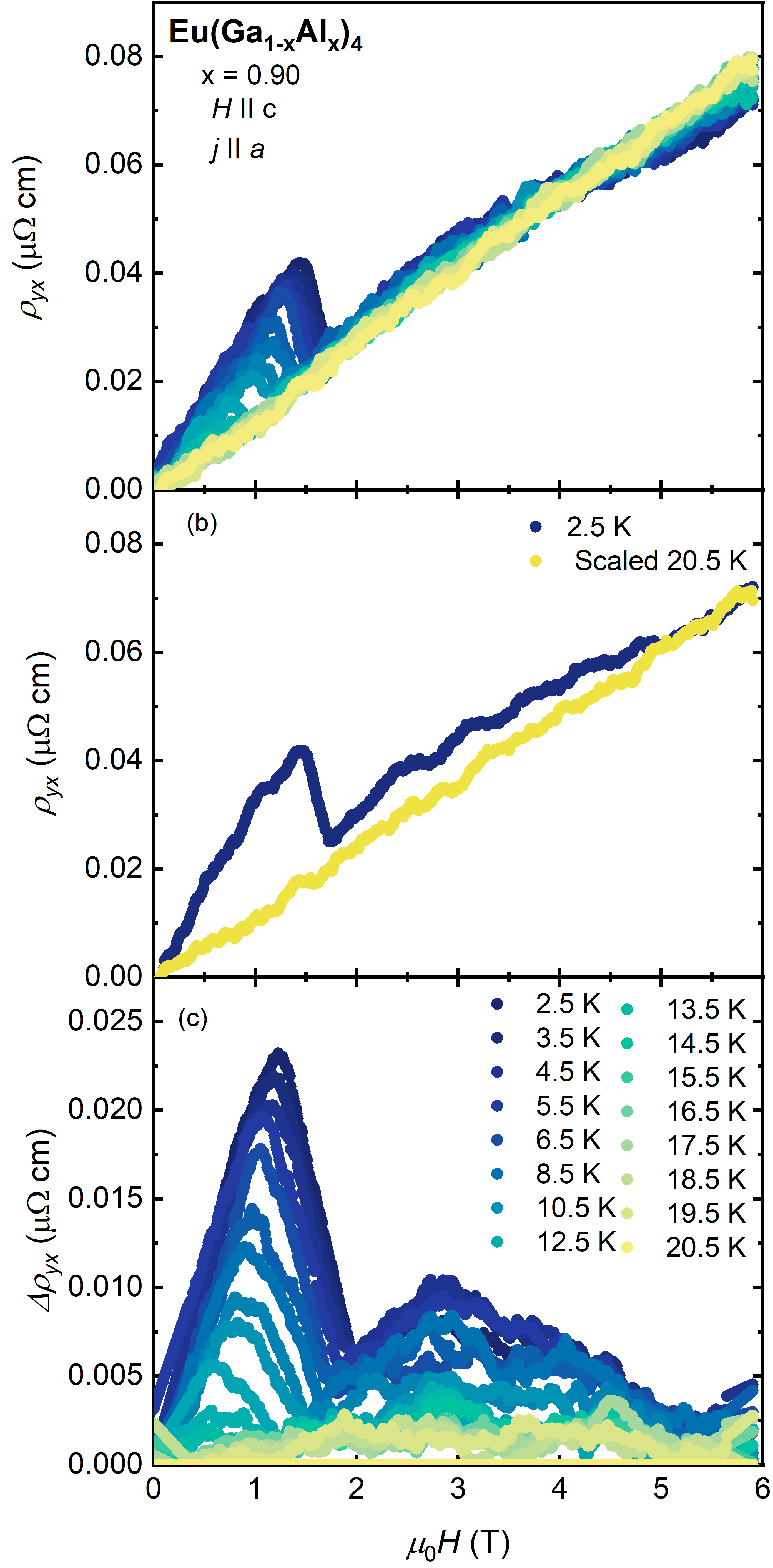}
\caption{\label{zerop9} (a) The Hall resistivity $\rho_{yx}$ measured with magnetic field $H~\parallel~c$ and current $j~\parallel~a$ for fields $0~\text{T}~\leq~\mu_0H~\leq~6~$T and temperatures $2.5~\text{K}~\leq~T~\leq20.5~$K for \EuGa~$x$~=~0.90.  (b)$\rho_{yx}$ at $T~=~2.5~$K (blue symbols) and $\rho_{yx}$ at $T~=~20.5~$K (yellow symbols) scaled to the high-field regime of of the $T~=~2.5~$K data. (c) The difference between the lower temperature $\rho_{yx}$ and scaled  $T~=~20.5~$K data, $\Delta\rho_{yx}$ for $2.5~\text{K}~\leq~T~\leq19.5~$K.}
\end{figure}

\begin{figure*}
\includegraphics[width=\textwidth]{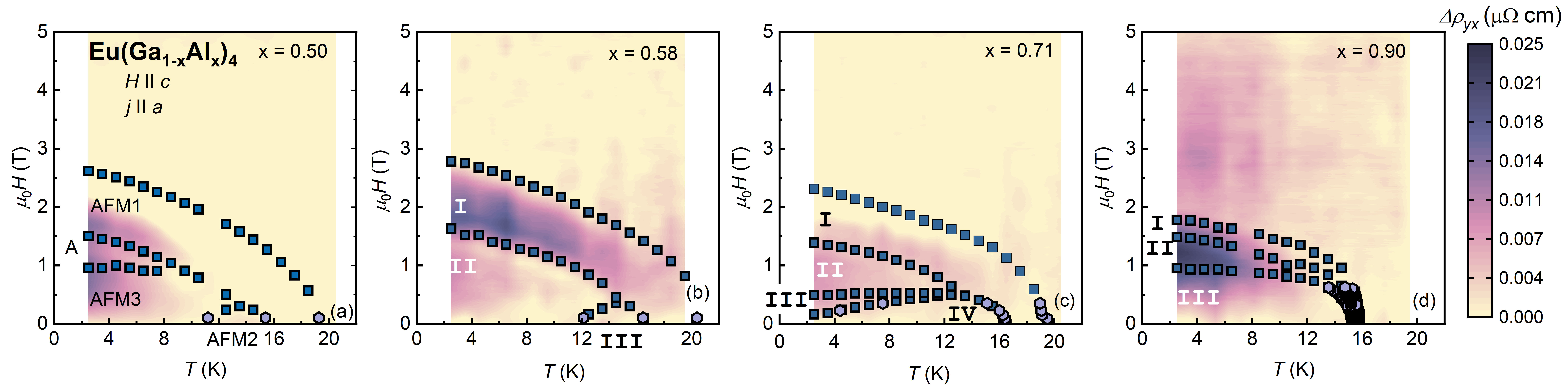}
\caption{\label{PhaseD} Magnetic field~-~temperature $H~-~T$ phase diagrams with magnetic field $H~\parallel~c$ for \EuGa~(a) $x$ = 0.50, (b) $x$ = 0.58, (c) $x$ = 0.71 and (d) $x$ = 0.90. Blue squares are determined from maxima in d$M$/d$H$ while purple hexagons are determined from d$(MT)$/d$T$. The color of the contour maps correspond to the contribution to the topological Hall resistivity $\Delta\rho_{yx}$.}
\end{figure*}

%To gain more quantitative information about the Hall effects in \EuGa, we perform a detailed analysis of $\rho_{yx}$.  Typically, 
The measured Hall resistivity $\rho_{yx}$ for a metallic magnet can have multiple contributions: 

\begin{equation}
    \rho_{yx} = R_0\mu_0H+\rho_{yx}^A+\Delta\rho_{yx}.
    \label{Eq:TotalHall}
\end{equation}

Here, the first term $R_0\mu_0H$ is the normal Hall resistivity arising from the Lorentz force the charge carriers experience as they traverse a perpendicular magnetic field $\mu_0H$, where $R_0$ is the Hall coefficient. When the system has one dominant carrier, the normal Hall effect is typically linear in field. 

The second term $\rho_{yx}^A$ is the anomalous Hall resistivity which scales with $M$. $\rho_{yx}^A$ can also have multiple contributions; however, depending on the magnitude of the conductivity $\sigma_{xx}~=~\rho_{xx}/(\rho_{xx}^2+\rho_{yx}^2)$, one contribution usually dominates. Empirically, for highly conductive systems \textit{i.e.} $\sigma_{xx}~>~10^6~(\Omega \text{cm})^{-1}$, the anomalous Hall conductivity $\sigma_{xy}^A~=~\rho_{yx}^A/(\rho_{xx}^2+\rho_{yx}^2)$ is dominated by skew-scattering and varies linearly with $\sigma_{xx}$ \cite{nagaosa2010anomalous}. Assuming $\rho_{yx}~<<~\rho_{xx}$, $\rho_{yx}^A~=~S_HM\rho_{xx}$, when skew scattering dominates, where $S_H$ is the anomalous Hall coefficient. 

The so-called intrinsic regime, where the AHE is dominated by the KL mechanism \cite{karplus1954hall}, is  empirically found to be in the range $10^4~(\Omega \text{cm})^{-1}~<~\sigma_{xx}~<~10^6~(\Omega \text{cm})^{-1}$ \cite{nagaosa2010anomalous}. In this regime, $\sigma_{xy}^A$ is roughly independent of $\sigma_{xx}$, or $\sigma_{xy}^A~\sim~constant$ \cite{nagaosa2010anomalous} and $\rho_{yx}^A~=~S_H'M\rho_{xx}^2$ \cite{lee2007hidden}. In the bad metal regime, $\sigma_{xx}~<~10^4~(\Omega \text{cm})^{-1}$, $\sigma_{yx}^A~\propto~\sigma_{xx}^n$ where $n~\sim 1.6$ \cite{nagaosa2010anomalous,onoda2006intrinsic}. All samples studied in this work display large conductivities  ($\sigma_{xx}~>~10^4~(\Omega \text{cm})^{-1}$), therefore the remainder of this manuscript is concerned with the highly conductive and intrinsic regimes. 

The third term in Eq. \ref{Eq:TotalHall}, $\Delta\rho_{yx}$, is the THE and is due to the non-zero SSC generated by non-coplanar spin textures. Such a contribution has been observed in both trivial non-coplanar spin textures \cite{wang2021field, ghimire2020competing}, as well as topological spin textures \cite{tokura2020magnetic,kurumaji2019skyrmion,hirschberger2019skyrmion,lee2009unusual}. 

To parse out the different contributions to the Hall effect in \EuGa, two different methods are used. The first method is valid when $\rho_{yx} << \rho_{xx}$ and assumes a single dominant carrier type. This method is applicable where both topological Hall and anomalous Hall contributions to $\rho_{yx}$ are expected, and when $\rho_{yx}(H > H_c)$ is linear.  Here, Eq.~\ref{Eq:TotalHall} can be written as 

\begin{equation}
\label{Eq:Skew}
    \rho_{yx} = R_0\mu_0H+S_H\rho_{xx}M + \Delta\rho_{yx}
\end{equation}
 
 \noindent assuming skew scattering as the dominant scattering mechanism, or 
 
 \begin{equation}
\label{Eq:instrinsic}
    \rho_{yx} = R_0\mu_0H+S_H'\rho_{xx}^2M + \Delta\rho_{yx}
\end{equation}

\noindent assuming the intrinsic KL-type scattering mechanism \cite{nagaosa2010anomalous}.

For fields greater than $\mu_0H_c$, $\Delta\rho_{yx}$ is necessarily zero since all spins are aligned in the SP state, and  Eqs.~\ref{Eq:Skew} and \ref{Eq:instrinsic} become

 \begin{equation}
\label{Eq:skewII}
    \frac{\rho_{yx}}{\mu_0H} = R_0+S_H\frac{\rho_{xx}M}{\mu_0H} 
\end{equation}

\noindent and

 \begin{equation}
\label{Eq:intrinsicII}
    \frac{\rho_{yx}}{\mu_0H} = R_0+S_H'\frac{\rho_{xx}^2M}{\mu_0H}, 
\end{equation}

\noindent respectively. $R_0$ and $S_H$ or $S_H'$ can thus be extracted as the intercept and slope of the corresponding line generated by plotting  $\rho_{yx}/\mu_0H$ \textit{vs.} $\rho_{xx}M/\mu_0H$ or \textit{vs.} $\rho_{xx}^2M/\mu_0H)$. $\Delta\rho_{yx}$ is then estimated below $\mu_0H_c$ as the difference

 \begin{equation}
\label{Eq:Deltarhoyx}
    \Delta\rho_{yx} = \rho_{yx}-R_0\mu_0H-\rho_{yx}^A.
\end{equation}

This analysis is demonstrated in Fig.~\ref{seventyone} for \EuGa~ with $x~=~0.71$, where the temperature dependence of $M(H)$, $\rho_{xx}(H)$ and $\rho_{yx}(H)$ are shown for $H~\parallel~c$ and $j~\parallel~a$ (panels a-c).  Fits of $\rho_{yx}(H)$ at $T~=~2.5~$K (blue) and  $T~=~20.5~$K (yellow) to  Eq.~\ref{Eq:skewII} (cyan line) and Eq.~\ref{Eq:intrinsicII} (pink line) are presented in Fig.~\ref{seventyone}d. For $T~=~2.5~$K, both equations capture the behavior of $\rho_{yx}(H)$ for  $\mu_0H~>~\mu_0H_c$, while for $\mu_0H~<~\mu_0H_c$, there are clear deviations of the measured $\rho_{yx}(H)$ compared to the fits, indicating an additional contribution $\Delta\rho_{yx}$.  When $T~=~20.5~$K, both  equations capture all features of $\rho_{yx}(H)$ such that $\Delta\rho_{yx}~=~0$.

The temperature and field dependence of $\Delta\rho_{yx}$ obtained after subtracting the the normal and anomalous contributions from the measured $\rho_{yx}$ (Eq.~\ref{Eq:Deltarhoyx}) are shown in Fig.~\ref{seventyone}e,f evaluated from fits to Eq.~\ref{Eq:skewII} and \ref{Eq:intrinsicII}, respectively. Qualitatively, both methods for estimating $\Delta\rho_{yx}$ result in similar temperature and field dependencies. However, Eq. \ref{Eq:intrinsicII} does a slightly better job of minimizing $\Delta\rho_{yx}$, consistent with Ref. \cite{moya2022incommensurate}. A similar analysis was done for \EuGa~ for x = 0.50 and 0.58 (the Supplementary Materials Fig.~\ref{THE_5} and \ref{THE_58}).

This method works for $x$ = 0.50, 0.58, and 0.71. However, a different method needs to be used to evaluate the THE for $x = 0.90$ because $\rho_{yx}$ above $\mu_0H_c$ shows a nonlinear field dependence (Fig.~\ref{zerop9}) for this composition, which requires a multi-band description for the normal Hall effect, instead of a single band analysis.

%, but the multi-band nature of the high field $\rho_{yx}$ (Fig.~\ref{HighField}) requires a different analysis. 
To extract $\Delta\rho_{yx}$ for \EuGa~ with $x$ = 0.90, we follow the procedure done in Ref.~\cite{suzuki2016large}. Assuming there is only a weak temperature dependence of $R_0$ and $\rho_{yx}^A$, the scaled $\rho_{yx}$ above $T_N$ is used to extract the THE at low temperatures. Here, the $T~=~20.5~$K (20.5~K$~>~T_N$) data is scaled to the lower temperature data, such that the high field data collapse onto each other. Fig.~\ref{zerop9}b  shows this analysis where the $T~=~20.5~$K $\rho_{yx}$ data (yellow) has been scaled to the $T~=~2.5~$K $\rho_{yx}$ data (blue).
$\Delta\rho_{yx}$ shown in Fig.~\ref{zerop9}c for all measured $T$ is obtained by subtracting the scaled high temperature from the lower temperature data, showing a clear THE. Such treatment precludes the analysis of $\rho_{yx}^A$.

%and present the result in Fig.~\ref{zerop9}. The temperature dependence of $\rho_{yx}(\mu_0H)$ is shown in Fig.~\ref{zerop9}a. To account for the weak temperature dependence of $R_0$ and $\rho_{yx}^A$,  In Fig.~\ref{zerop9}b we show this analysis where the $T~=~20.5~$K data (yellow) has been scaled to the $T~=~2.5~$K data (blue).  

As a result of this analysis, a contour map of $\Delta\rho_{yx}$ is produced, shown in Fig.~\ref{PhaseD}(a-d) for \EuGa~ with $x$ = 0.50, 0.58, 0.71 and 0.90. The $H-T$ phase diagrams determined from isothermal magnetization (blue squares) and temperature-dependent magnetic susceptibility measurements (purple hexagons)\cite{moya2022incommensurate,stavinoha2018charge} are shown together with the contour plots .
%We now focus our attention to the relationship between $\Delta\rho_{yx}$ and the $H \parallel c$ $H-T$ phase diagrams shown in Fig.~\ref{PhaseD}(a-d) for \EuGa~x = 0.50, 0.58, 0.71 and 0.90, respectively. In Fig.~\ref{PhaseD}, the $H-T$ phase diagrams (symbols) are constructed , where the , while $\Delta\rho_{yx}$ is plotted as a contour map using data from Supplemental \textcolor{red}{find fig}f,\textcolor{red}{find fig}f, Fig.~\ref{seventyone}f and Fig.~\ref{zerop9}c for compositions x = 0.50, 0.58, 0.71 and 0.90, respectively.  

%Qualitatively, the  $H-T$ phase diagrams (Fig.~\ref{PhaseD}(a-d)) do not change in a simple fashion. However, 
Overall, non-zero $\Delta\rho_{yx}$ can be observed in certain phase region for all four compositions. For $x = 0.5$ (Fig.~\ref{PhaseD}a), both the $H-T$ phase diagram and $\Delta\rho_{yx}$ map are consistent with previous results \cite{moya2022incommensurate}. Here, the AFM1 phase for H = 0 was determined to be an incommensurate helical structure with a magnetic propagation vector $q_{mag}$ along the crystallographic $a$ axis, and magnetic moments rotating in in the $ab$ plane. The ground state, AFM3, was determined to be a cycloidal state again with $q_{mag}$ along the crystollographic $a$ axis, but with moments rotating in the $bc$ plane. The phase AFM2 that separates AFM1 and AFM3 on cooling was determined to have mixed magnetic propagation vectors. With the application of magnetic field $H \parallel c$, AFM3 and AFM1 are separated by the intermediate field phase (A phase), where  $\Delta\rho_{yx}$ is centered around, indicative of a topological spin texture or a more generic non-coplanar spin texture.

\begin{figure}
\includegraphics[width=\columnwidth]{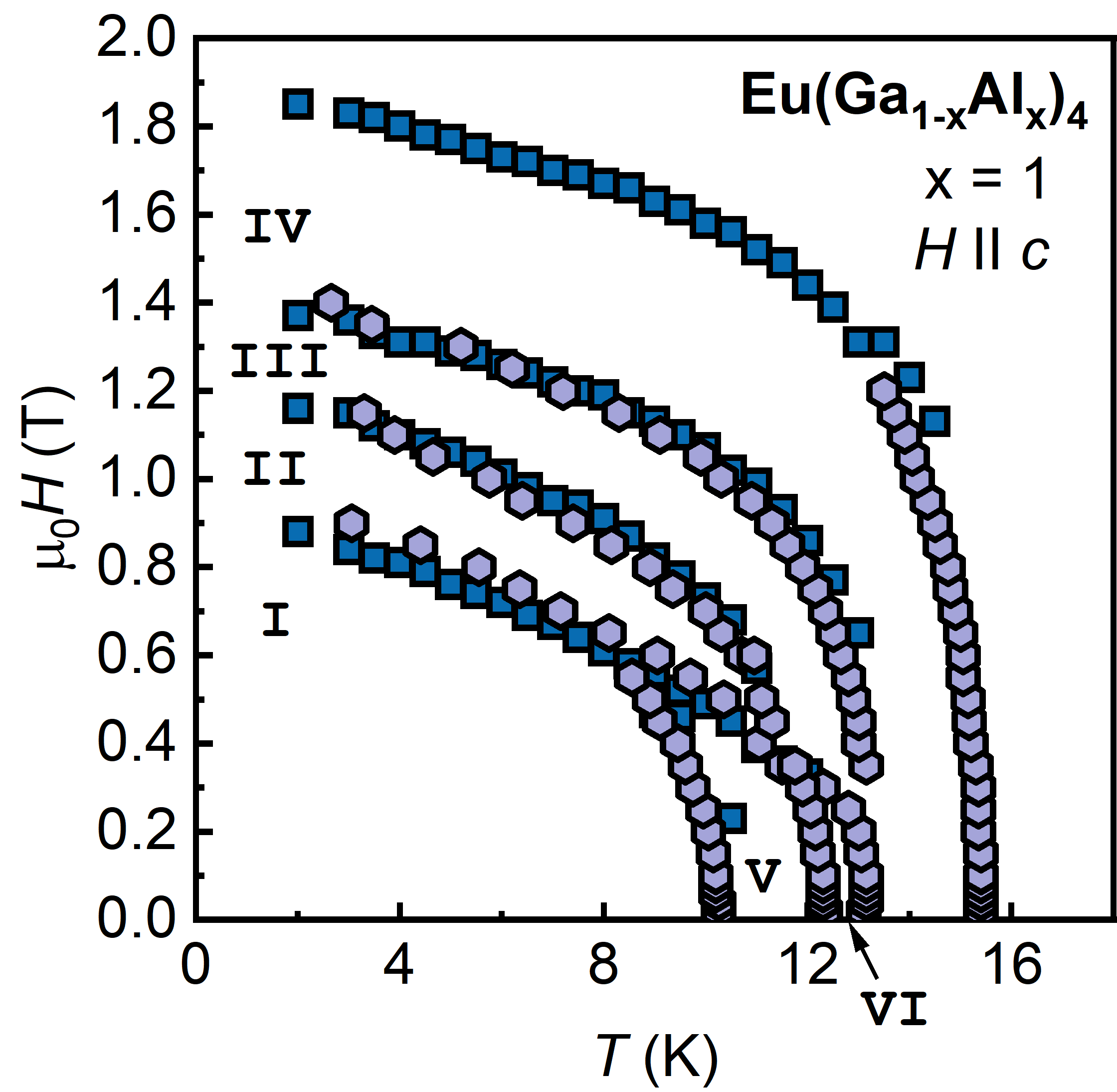}
\caption{\label{EuAl4} Magnetic field - temperature phase diagram of EuAl$_4$ (\EuGa~ $x$ = 1) for $H \parallel c$. The phase boundaries are determined by maxima in d$M$/d$H$ (blue squares) and d$(MT)$/d$T$ (purple hexagons). }
\end{figure}

For \EuGa with $x$ = 0.58, 0.71 and 0.90 (Fig.~\ref{PhaseD}b-d), the maximum THE appears in the regions I, II, and II, respectively.  We focus on \EuGa~ $x$= 0.9, the compound closest in composition to the confirmed skyrmion host EuAl$_4$ \cite{takagi2022square}. Despite the existence of the THE in both compounds (the THE in EuAl$_4$ was reported in Ref. \cite{shang2021anomalous}), the phase diagrams are strikingly different. For comparison, the $H~-~T$ phase diagram ($H \parallel c$) determined by magnetization measurements for EuAl$_4$ is presented in Fig.~\ref{EuAl4} which is consistent with previous reports \cite{takagi2022square}. EuAl$_4$ possesses four different magnetically ordered phases on zero-field cooling, labeled IV, VI, V and I in Fig. \ref{EuAl4}. This is in contrast to \EuGa~ with $x$ = 0.9, which only has two phases below $T_N$, labeled I and III in Fig. \ref{PhaseD}d. Also, with the application of magnetic field $H \parallel c$, the zero-field phases in EuAl$_4$ (Fig.~\ref{EuAl4}) are separated by two additional phases, phase II (rhombic skyrmion lattice) and III (square skyrmion lattice)  \cite{takagi2022square}, compared to only one intermediate field state in \EuGa~ $x$= 0.9 (phase II in Fig.~\ref{PhaseD}d) where the THE reaches a maximum.  The different $H -T$ phase diagrams reflect the ability to fine-tune the magnetic interactions in the two compounds using chemical substitution, possibly favoring one type of skyrmion lattice over the other. 

The Dzyaloshinskii-Moriya (DM) interaction \cite{dzyaloshinsky1958thermodynamic, moriya1960new} has been emphasized as a key ingredient in stabilizing topological spin textures in non-centrosymetric crystals. However, the \EuGa~ crystal structure is centrosymmetric and therefore DM interactions should be absent \cite{tokura2020magnetic}. Instead, recent theoretical studies \cite{ozawa2017zero,hayami2017effective,wang2020skyrmion,hayami2021square} suggested that the interplay between the Ruderman-Kittel-Kasuya-Yosida (RKKY) interaction \cite{ruderman1954indirect,kasuya1956theory, yosida1957magnetic} and the four-spin interaction could stabilize topological spin-textures in metallic centrosymmetric magnets.  In these theories \cite{ozawa2017zero,hayami2017effective,wang2020skyrmion,hayami2021square}, the RKKY interaction stabilizes $q_{mag}$ set by the Fermi surface nesting vector, while the four-spin interaction, which depends on the in-plane bond-dependent anisotropy and easy axis anisotropy, increases the propensity towards multi-$q$ order. Being able to tune these interactions in centrosymmetric compounds offers the opportunity for tunable topological spin textures, which is in contrast to their non-centrosymmetric counter parts, where the type of topological spin texture is set by the crystalline symmetry \cite{tokura2020magnetic}. Future experimental and theoretical works on \EuGa~ will elucidate the influence of chemical substitution on these tuning parameters, which may, in turn, enable targeted engineering of topological spin textures.

%These results in conjunction with  in the $H-T$ phase space where the THE is maximum \cite{shang2021anomalous}, demonstrate that it is highly likely there are also topological spin textures in \EuGa~for x $\geq$ 0.5.  

%In the field-induced spin-polarized state, the behavior of $\Delta\rho_{yx}$ stands out for \EuGa~x = 0.90 (Fig.~\ref{PhaseD}d), compared to x = 0.50, 0.58 and 0.71 (Fig.~\ref{PhaseD}a-c).  For \EuGa~x = 0.90,  $\Delta\rho_{yx}$ is non-zero above $\mu_0H_c$ at low temperatures, compared to zero for x = 0.50, 0.58 and 0.71. A similar feature has been observed in thin films of skyrmion host MnSi, where it is speculated that fluctuations of the non-coplanar spin texture above $\mu_0H_C$  can generate a large contribution to the Hall signal \cite{fujishiro2021giant}.

\subsection{Anomalous Hall effect (AHE)}

\begin{figure}
\includegraphics[width=\columnwidth]{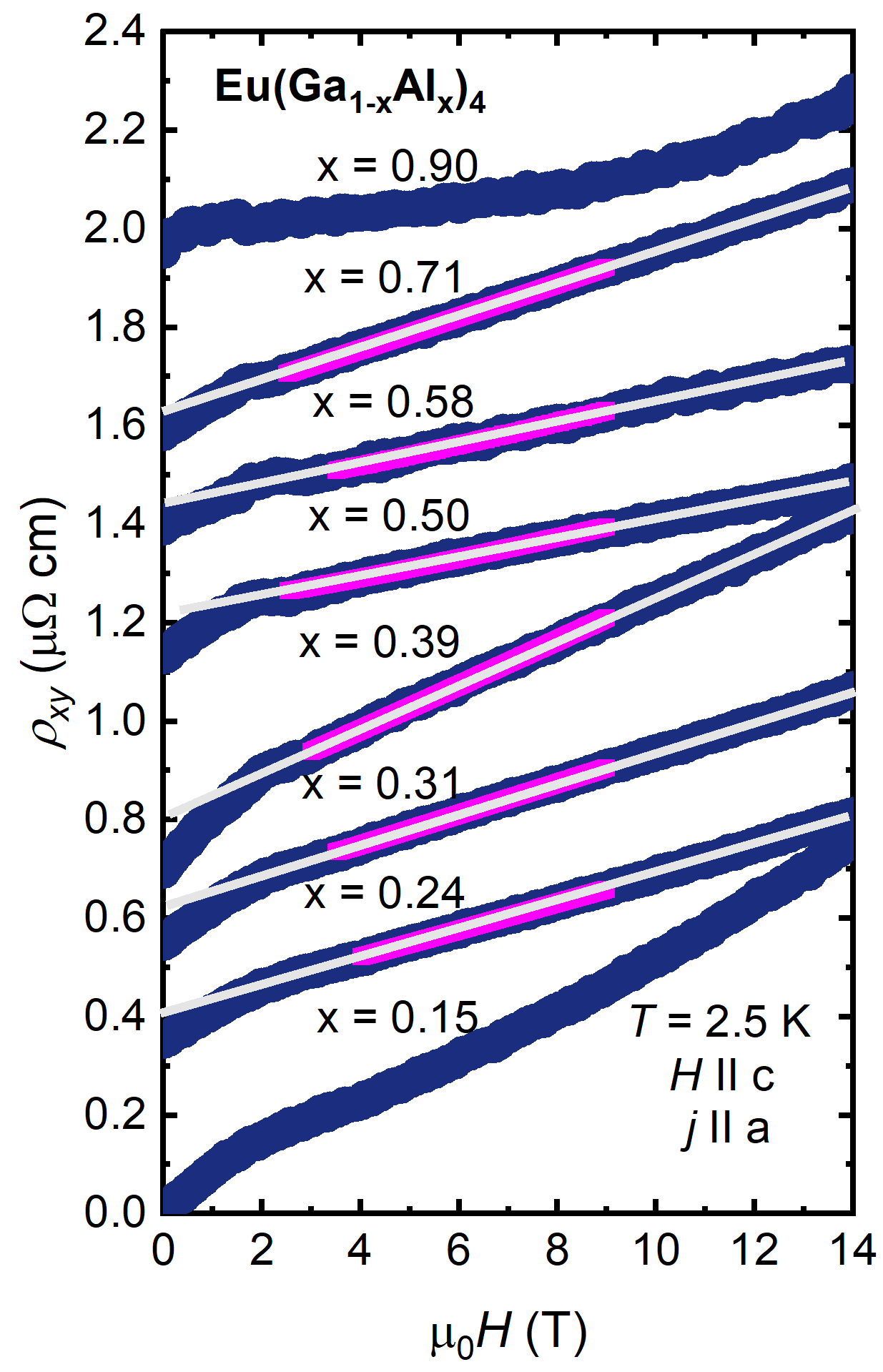}
\caption{\label{HighField} Magnetic field dependence of the Hall resistivity $\rho_{yx}$ (blue symbols) for \EuGa~ ($x$ = 0.15, 0.24, 0.31, 0.39, 0.50, 0.58, 0.71, 0.90) measured at $T = 2.5$ K with $H \parallel c$ and $j \parallel a$ for $0~\text{T} \leq \mu_0H \leq 14~\text{T}$. Data are offset such that $\rho_{yx}(H = 0) = 0$. Pink lines are fits to Eq.~ \ref{Eq:intrinsicII} in the field-polarized state up to $\mu_0H = 9$ T, the highest field we are able to measure magnetization. The grey lines are linear extrapolations of the pink lines.}
\end{figure}

\begin{figure}
\includegraphics[width=\columnwidth]{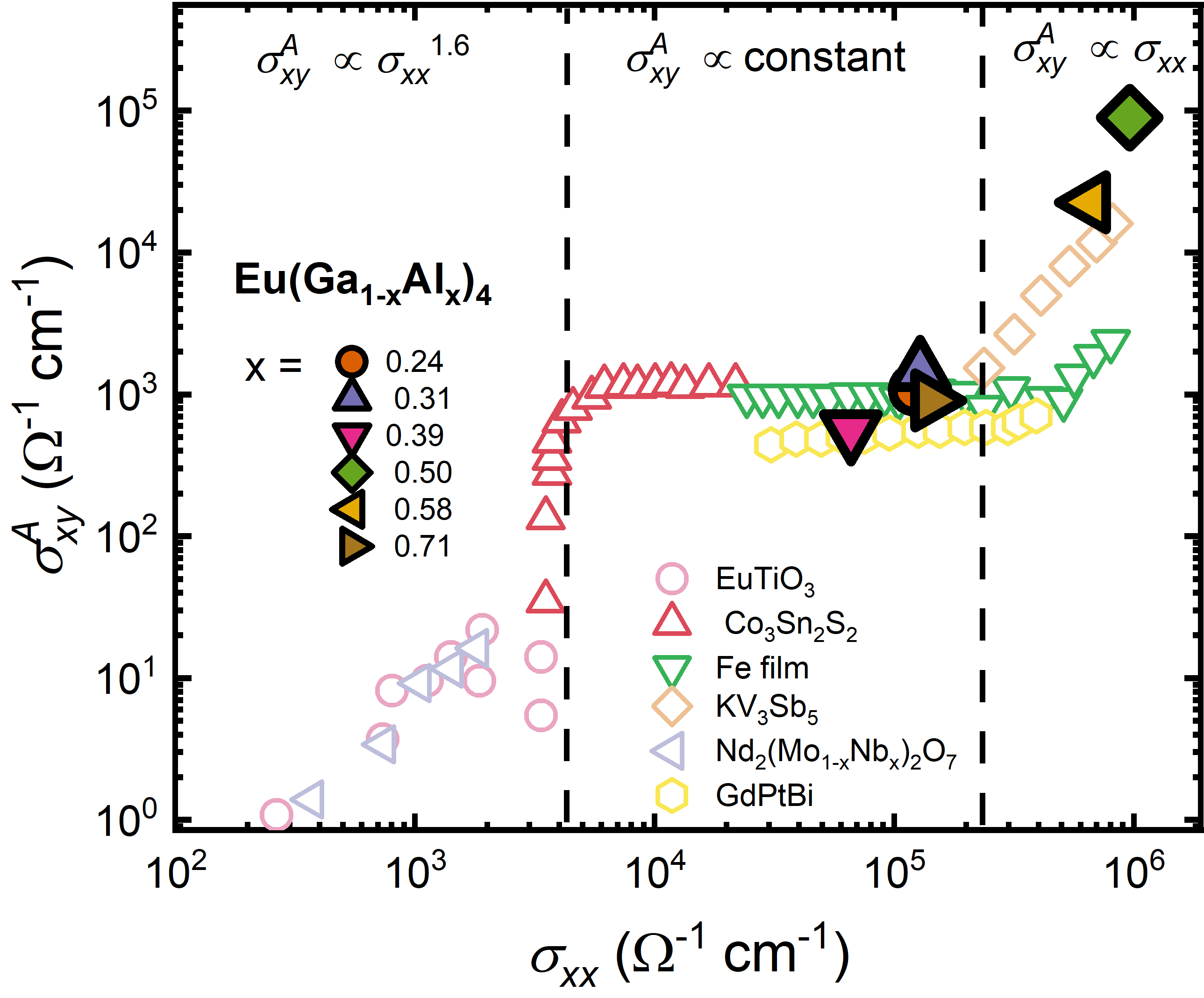}
\caption{\label{sigmaxy} The anomalous Hall conductivity $\sigma_{xy}^A$ plotted as a function of longitudinal conductivity $\sigma_{xx}$ for \EuGa~$x$~=~0.24, 0.31, 0.39, 0.50, 0.58 and 0.71 extracted from the field polarized regime ($\mu_0H = 9$T)  at temperature $T~=~2.5~$K with fits to  Eq.~\ref{Eq:intrinsicII} (large, closed symbols) compared to other metallic magnets (small open symbols) taken from Ref.~\cite{takahashi2018anomalous,liu2018giant,miyasato2007crossover,yang2020giant,iguchi2007scaling,suzuki2016large}.}
\end{figure}

\begin{figure}
\includegraphics[width=\columnwidth]{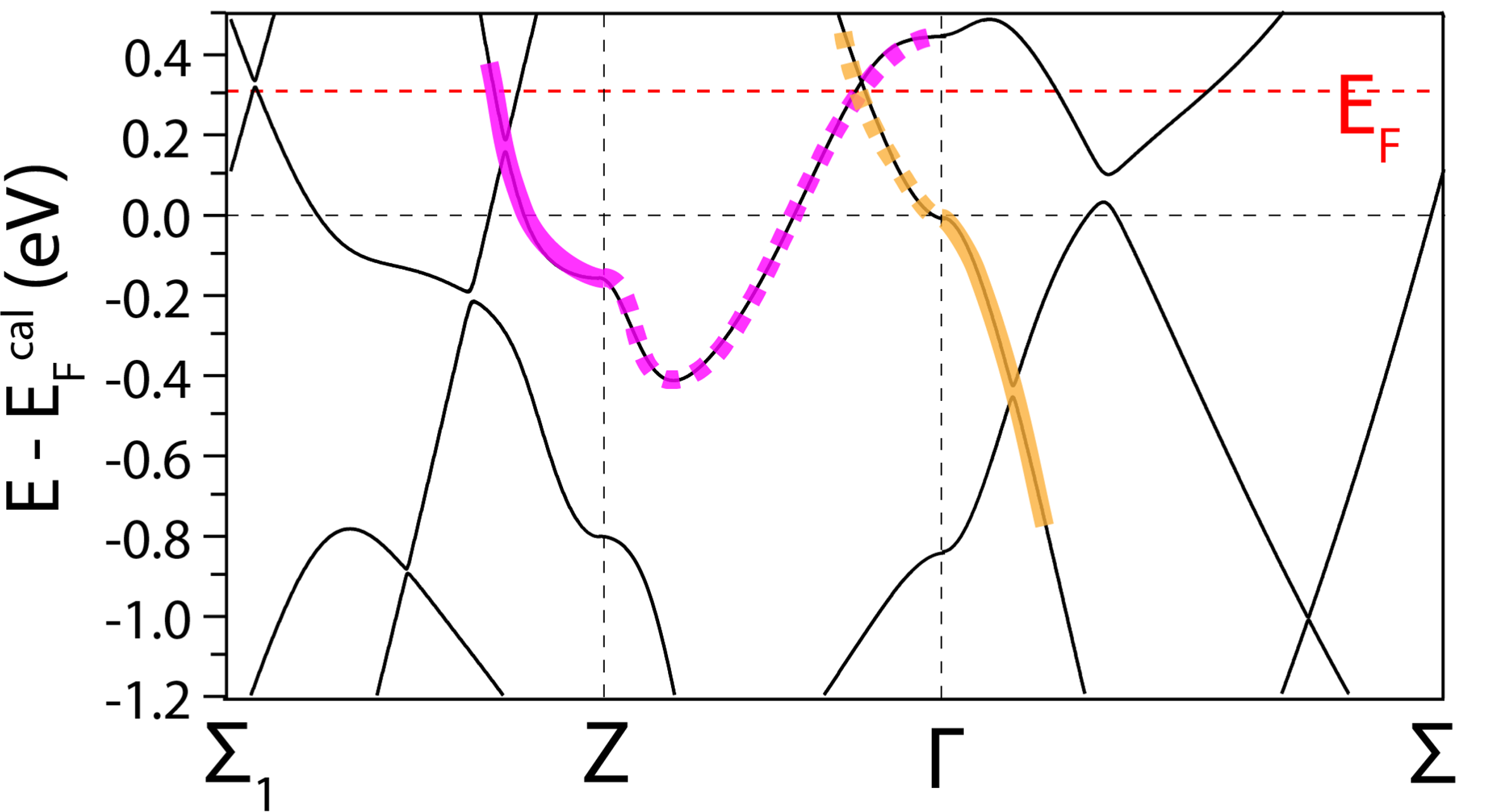}
\caption{\label{ARPES}  The electronic structure calculated along certain high symmetry lines for \EuGa, $x$ = 0.5, corresponding to the ordered structure EuGa$_2$Al$_2$, in the paramagnetic state. The yellow and magenta lines mark out the bands that form a Dirac point. The red dashed line indicates the Fermi level determined by the ARPES experiments \cite{Jaime2023}.}
\end{figure}

We now turn to the AHE in the SP state above $\mu_0H_c$ (where $M$ is fully saturated at $T = 2.5$ K). Extended measurements of $\rho_{yx}(H)$, $\rho_{xx}(H)$, and $M(H)$ up to $\mu_0H$ = 9 T are shown in Supplementary Materials Fig. \ref{two59T}.  Eq.~\ref{Eq:intrinsicII} is used to extract the anomalous Hall resistivity as $\rho_{yx}^A = S_H'\frac{\rho_{xx}^2M}{\mu_0H}$. Such an analysis is valid only for the single-band Hall effect so the analysis is restricted to \EuGa~ with $0.24~\leq~~x~\leq~0.71$.  Fits to the $T = 2.5$ K data are shown in Fig.~\ref{HighField} as pink lines for $\mu_0H_c \leq \mu_0H \leq 9$ T. Next,  $\rho_{yx}^A$ is converted to anomalous Hall conductivity $\sigma_{xy}^A$ using the tensor relation $\sigma_{xy}^A~=~\rho_{yx}^A/(\rho_{xx}^2+\rho_{yx}^2)$. In Fig.~\ref{sigmaxy} $\sigma_{xy}^A(\mu_0H = 9 \text{T}, T = 2.5 \text{K})$ is plotted for each composition compared to $\sigma_{xx}(\mu_0H = 0~\text{T}, T = 2.5~\text{K})$, which serves as a measure of disorder and naturally varies with $x$.

% Having measured $\rho_{yx}(\mu_0H)$, $\rho_{xx}(\mu_0H)$, and $M(\mu_0H)$ as shown in Fig.~\ref{HighField}a-c for \EuGa~ $x$ = 0.71 and Supplementary \textcolor{red}{insert figs} for other compositions, we use Eq.~\ref{Eq:intrinsicII} to extract the anomalous Hall resistivity as $\rho_{yx}^A = S_H'\frac{\rho_{xx}^2M}{\mu_0H}$. We reiterate that such an analysis is valid only for the single-band Hall effect so we restrict ourselves to \EuGa~$x$ = 0.24, 0.31, 0.39, 0.50, 0.58 and 0.71.  Fits to the $T = 2.5$ K data are shown in Fig.~\ref{HighField} as pink lines for $\mu_0H_c \leq \mu_0H \leq 9$ T, where we are limited to 9 T as this is the highest field we can measure $M$ in our instrument.  Next, we convert $\rho_{yx}^A$ to the anomalous Hall conductivity $\sigma_{xy}^A$ using the tensor relation $\sigma_{xy}^A~=~\rho_{yx}^A/(\rho_{xx}^2+\rho_{yx}^2)$. In Fig.~\ref{sigmaxy} we plot $\sigma_{xy}^A(\mu_0H = 9 \text{T}, T = 2.5 \text{K})$ for each composition compared to $\sigma_{xx}(\mu_0H = 0~\text{T}, T = 2.5~\text{K})$, which serves as a measure of disorder and naturally various with $x$.

For $x = 0.24,~0.31,~0.39, \text{and}~0.71$,  $\sigma_{xy}^A$ is nearly independent of $\sigma_{xx}$, consistent with the intrinsic KL mechanism. The resulting $\sigma_{xy}^A~=~600-1000~(\Omega~\text{cm})^{-1}$, is also comparable with the theoretical limit of $10^2\sim10^3~(\Omega \text{cm})^{-1}$ when $\sigma_{xy}^A$ is generated by reciprocal-space Berry curvature \cite{fujishiro2021giant}. The compounds with $x$ = 0.50 and 0.58 are much more conductive, entering the regime where $\sigma_{xy}^A$ is dominated by the skew-scattering mechanism. For these compositions, $\sigma_{xy}^A$ increases with increasing $\sigma_{xx}$, which appears to be in line with the skew scattering mechanism. However, the Hall angle $\Theta_H~=~\tan^{-1}(\sigma_{xy}^A/\sigma_{xx})$ of typical metals where skew scattering dominates is usually less than 1\% \cite{fujishiro2021giant}. $\Theta_H$ measured for \EuGa~with \textit{x} = 0.50 is $\sim$~7\% indicating that another mechanism (possibly intrinsic KL) is also contributing to the large $\sigma_{xy}^A$, a consequence of reciprocal space Berry curvature.
 
% Since $\sigma_{xy}^A$ is quite large (larger than those of KV3Sb5 and Fe film in the same regime in Fig.~\ref{sigmaxy}), the resulting the Hall angle $\Theta_H~=~\tan^{-1}(\sigma_{xy}^A/\sigma_{xx})$ is also quite large. For x = 0.50 an d 0.58, $\Theta_H$ is determined to be xx and xx, respectively. For reference, $\Theta_H$ of typical metals where skew-scattering dominates is usually less than 1\% \cite{fujishiro2021giant}. Such discrepancy suggests additional mechanisms contributing to $\sigma_{xy}^A$, such as xx and xx (maybe japanese paper on the giant anomalous one, and jeff's paper).  

%is supported evidence by the increase in $\sigma_{xy}^A$ as $\sigma_{xx}$ is increased from x = 0.58 to 0.50. However, 
%the Hall angle $\Theta_H~=~\tan^{-1}(\sigma_{xy}^A/\sigma_{xx})$ of typical metals where skew-scattering dominates is usually less than 1\% \cite{fujishiro2021giant}. $\Theta_H$ measured for \EuGa~x = 0.50 is $\sim$~7\% indicating that another mechanism  like the intrinsic KL mechanism, a consequence of reciprocal space Berry curvature, is contributing to the large $\sigma_{xy}^A$.

%\subsection{Observation of a Dirac point near the Fermi level}
%\label{DP}

First-principles calculations are used to further study the topological electronic structure, which is the origin of the reciprocal-space Berry curvature for the \EuGa~ series.  The band structure of EuGa$_2$Al$_2$ in the paramagnetic state obtained by density functional theory (DFT) calculations is presented in Fig~\ref{ARPES}, which shows a Dirac point along the $\Gamma-Z$ path. The Dirac point is protected by the four-fold rotational symmetry with respect to the $z$ axis in this tetragonal system: the two bands highlighted with magenta and yellow lines have different irreducible representations of the $C_{4v}$ point group and cross each other along $\Gamma-Z$ without opening a gap when spin-orbit coupling is considered. Note that this type of Dirac crossing is sometimes also called semi-Dirac or anisotropic Dirac crossing because the band dispersion is linear along one axis ($k_z$), while being quadratic along other axes ($k_x$ and $k_y$). Such a crossing has been discussed in the study on the non-magnetic isostructural analogue BaAl$_4$ \cite{wang2020crystalline}. However, the Dirac crossing is located $\sim$0.4 eV above the Fermi level in BaAl$_4$.

The existence of the Dirac point in EuGa$_2$Al$_2$ is verified by ARPES measurements \cite{Jaime2023}. In the Brillouin zone (BZ) $Z$ plane, there is a hole band (yellow dashed line) and an electron band (magenta dashed line) which intersect each other around the BZ center.  The ARPES measurements \cite{Jaime2023} agree well with the DFT calculations and confirm the existence of the Dirac point along the $\Gamma-Z$ path. More importantly, the Dirac point in EuGa$_2$Al$_2$ is found to be located close to the Fermi level by angle-resolved photomession spectroscopy measurements \cite{Jaime2023}, in contrast to the case of BaAl$_4$ \cite{wang2020crystalline}. In the SP state, the spin-degeneracy of the bands will be lifted, which could lead to the formation of Weyl points. The mechanism is similar to that of GdPtBi \cite{hirschberger2016chiral,shekhar2018anomalous,cano2017chiral}, where the gapless quadratic band touching in the electronic structure sets the stage for the field induced Weyl physics and the AHE.

Finally, it is noted that the energy of the Dirac point in \EuGa~ series may slightly vary as Al/Ga ratio changes. In EuGa$_4$, for example, the Dirac point was determined to be above the Fermi level \cite{Kevin2021}. The origin is likely related to the difference in electronegativity between Al and Ga. Consequently, a change in Al/Ga ratio in \EuGa~ also provides an opportunity to change the intrinsic anomalous Hall conductivity.

%Generically speaking, when the time-reversal symmetry in Dirac semimetal is broken, the bands become spin split resulting in Weyl nodes \cite{armitage2018weyl}.  This system has non-zero Berry curvature where the resulting $\sigma_{xy}^A$ is proportional to the spacing of the Weyl nodes in momentum space\cite{armitage2018weyl}.
%Therefore, together our electrical transport and ARPES data suggests the large $\sigma_{xy}$ we observe in \EuGa~ is a consequence of Weyl nodes which should appear in the spin-polarized state when the TRS is broken. We note that Al and Ga are isovalent, such that we would naively expect little to no change in the chemical potential across the \EuGa~ series.  However, the chemical potential measured by ARPES in EuGa$_2$Al$_2$ is $\sim0.4$ eV higher than that measured in EuAl$_4$ \cite{Kevin2021}.  Further ARPES studies would be needed, gain an understanding of how the chemical potential changes in the \EuGa~ series, but the interpretation of $\sigma_{xy}^A$ is still valid since $\sigma_{xy}^A$ in Weyl-semimetals has been shown to be relatively robust to changes in the chemical potential \cite{burkov2014anomalous,armitage2018weyl}.

%compared to at the Fermi energy in EuGa$_2$Al$_2$

\section{Conclusions}

In conclusion, magnetotransport measurements on \EuGa~ single crystals with $0.15~\leq~x~\leq~0.9$ reveal evidence for a THE for $x~\geq~0.5$, pointing to the existence of non-coplanar spin textures. For intermediate composition $0.24~<~x~<~0.39$, the Hall resistivity varies smoothly with magnetic field, although multiple magnetic ordered phases are clearly revealed in the isothermal magnetization and longitudinal resistivity measurements. Our measurements suggest a weak THE contribution, if any of the phase regimes host topological spin textures. At the lowest doping $x$ = 0.15, only one magnetic ordered phase can be observed before reaching the SP state, similar to the end compound EuGa$_4$. Therefore, no topological spin texture is expected. In the SP state,  evidence for a large AHE in \EuGa~is found, which is attributed to the appearance of Weyl nodes, generated after spin-splitting of a Dirac point.  The existence of the Dirac point near the Fermi level in the paramagnetic state in EuGa$_2$Al$_2$ is confirmed by DFT calculations which has been verified by ARPES measurements.  Therefore \EuGa~  for $x~\geq~0.5$ is established as a rare platform to study the field-tunable phenomena associated with reciprocal- and real-space topology.

\begin{acknowledgments}
YG, SL and EM acknowledge partial support from AFOSR Grant no. FA9550-21-1-0343. JMM has been supported by the NSF Graduate Research Fellowship Program under Grant no. DGE-1842494 and National Science Foundation (NSF) DMR Grant no. 1903741. KA and EM have been partially supported by the Robert A. Welch Foundation grant C-2114.  The use of the EPMA facility at the Department of Earth Science, Rice University, Houston, Texas, is kindly acknowledged. The authors thank Gelu Costin for help with EPMA measurements.

\end{acknowledgments}

\bibliography{AHE}

\begin{thebibliography}{78}%
\makeatletter
\providecommand \@ifxundefined [1]{%
 \@ifx{#1\undefined}
}%
\providecommand \@ifnum [1]{%
 \ifnum #1\expandafter \@firstoftwo
 \else \expandafter \@secondoftwo
 \fi
}%
\providecommand \@ifx [1]{%
 \ifx #1\expandafter \@firstoftwo
 \else \expandafter \@secondoftwo
 \fi
}%
\providecommand \natexlab [1]{#1}%
\providecommand \enquote  [1]{``#1''}%
\providecommand \bibnamefont  [1]{#1}%
\providecommand \bibfnamefont [1]{#1}%
\providecommand \citenamefont [1]{#1}%
\providecommand \href@noop [0]{\@secondoftwo}%
\providecommand \href [0]{\begingroup \@sanitize@url \@href}%
\providecommand \@href[1]{\@@startlink{#1}\@@href}%
\providecommand \@@href[1]{\endgroup#1\@@endlink}%
\providecommand \@sanitize@url [0]{\catcode `\\12\catcode `\$12\catcode
  `\&12\catcode `\#12\catcode `\^12\catcode `\_12\catcode `\%12\relax}%
\providecommand \@@startlink[1]{}%
\providecommand \@@endlink[0]{}%
\providecommand \url  [0]{\begingroup\@sanitize@url \@url }%
\providecommand \@url [1]{\endgroup\@href {#1}{\urlprefix }}%
\providecommand \urlprefix  [0]{URL }%
\providecommand \Eprint [0]{\href }%
\providecommand \doibase [0]{http://dx.doi.org/}%
\providecommand \selectlanguage [0]{\@gobble}%
\providecommand \bibinfo  [0]{\@secondoftwo}%
\providecommand \bibfield  [0]{\@secondoftwo}%
\providecommand \translation [1]{[#1]}%
\providecommand \BibitemOpen [0]{}%
\providecommand \bibitemStop [0]{}%
\providecommand \bibitemNoStop [0]{.\EOS\space}%
\providecommand \EOS [0]{\spacefactor3000\relax}%
\providecommand \BibitemShut  [1]{\csname bibitem#1\endcsname}%
\let\auto@bib@innerbib\@empty
%</preamble>
\bibitem [{\citenamefont {Haldane}(1988)}]{haldane1988model}%
  \BibitemOpen
  \bibfield  {author} {\bibinfo {author} {\bibfnamefont {F.~D.~M.}\
  \bibnamefont {Haldane}},\ }\href@noop {} {\bibfield  {journal} {\bibinfo
  {journal} {Physical review letters}\ }\textbf {\bibinfo {volume} {61}},\
  \bibinfo {pages} {2015} (\bibinfo {year} {1988})}\BibitemShut {NoStop}%
\bibitem [{\citenamefont {Liu}\ \emph {et~al.}(2016)\citenamefont {Liu},
  \citenamefont {Zhang},\ and\ \citenamefont {Qi}}]{liu2016quantum}%
  \BibitemOpen
  \bibfield  {author} {\bibinfo {author} {\bibfnamefont {C.-X.}\ \bibnamefont
  {Liu}}, \bibinfo {author} {\bibfnamefont {S.-C.}\ \bibnamefont {Zhang}}, \
  and\ \bibinfo {author} {\bibfnamefont {X.-L.}\ \bibnamefont {Qi}},\
  }\href@noop {} {\bibfield  {journal} {\bibinfo  {journal} {Annual Review of
  Condensed Matter Physics}\ }\textbf {\bibinfo {volume} {7}},\ \bibinfo
  {pages} {301} (\bibinfo {year} {2016})}\BibitemShut {NoStop}%
\bibitem [{\citenamefont {Berry}(1984)}]{berry1984quantal}%
  \BibitemOpen
  \bibfield  {author} {\bibinfo {author} {\bibfnamefont {M.~V.}\ \bibnamefont
  {Berry}},\ }\href@noop {} {\bibfield  {journal} {\bibinfo  {journal}
  {Proceedings of the Royal Society of London. A. Mathematical and Physical
  Sciences}\ }\textbf {\bibinfo {volume} {392}},\ \bibinfo {pages} {45}
  (\bibinfo {year} {1984})}\BibitemShut {NoStop}%
\bibitem [{\citenamefont {Chang}\ \emph {et~al.}(2013)\citenamefont {Chang},
  \citenamefont {Zhang}, \citenamefont {Feng}, \citenamefont {Shen},
  \citenamefont {Zhang}, \citenamefont {Guo}, \citenamefont {Li}, \citenamefont
  {Ou}, \citenamefont {Wei}, \citenamefont {Wang} \emph
  {et~al.}}]{chang2013experimental}%
  \BibitemOpen
  \bibfield  {author} {\bibinfo {author} {\bibfnamefont {C.-Z.}\ \bibnamefont
  {Chang}}, \bibinfo {author} {\bibfnamefont {J.}~\bibnamefont {Zhang}},
  \bibinfo {author} {\bibfnamefont {X.}~\bibnamefont {Feng}}, \bibinfo {author}
  {\bibfnamefont {J.}~\bibnamefont {Shen}}, \bibinfo {author} {\bibfnamefont
  {Z.}~\bibnamefont {Zhang}}, \bibinfo {author} {\bibfnamefont
  {M.}~\bibnamefont {Guo}}, \bibinfo {author} {\bibfnamefont {K.}~\bibnamefont
  {Li}}, \bibinfo {author} {\bibfnamefont {Y.}~\bibnamefont {Ou}}, \bibinfo
  {author} {\bibfnamefont {P.}~\bibnamefont {Wei}}, \bibinfo {author}
  {\bibfnamefont {L.-L.}\ \bibnamefont {Wang}},  \emph {et~al.},\ }\href@noop
  {} {\bibfield  {journal} {\bibinfo  {journal} {Science}\ }\textbf {\bibinfo
  {volume} {340}},\ \bibinfo {pages} {167} (\bibinfo {year}
  {2013})}\BibitemShut {NoStop}%
\bibitem [{\citenamefont {Tokura}\ and\ \citenamefont
  {Kanazawa}(2020)}]{tokura2020magnetic}%
  \BibitemOpen
  \bibfield  {author} {\bibinfo {author} {\bibfnamefont {Y.}~\bibnamefont
  {Tokura}}\ and\ \bibinfo {author} {\bibfnamefont {N.}~\bibnamefont
  {Kanazawa}},\ }\href@noop {} {\bibfield  {journal} {\bibinfo  {journal}
  {Chemical Reviews}\ }\textbf {\bibinfo {volume} {121}},\ \bibinfo {pages}
  {2857} (\bibinfo {year} {2020})}\BibitemShut {NoStop}%
\bibitem [{\citenamefont {Fert}\ \emph {et~al.}(2013)\citenamefont {Fert},
  \citenamefont {Cros},\ and\ \citenamefont {Sampaio}}]{fert2013skyrmions}%
  \BibitemOpen
  \bibfield  {author} {\bibinfo {author} {\bibfnamefont {A.}~\bibnamefont
  {Fert}}, \bibinfo {author} {\bibfnamefont {V.}~\bibnamefont {Cros}}, \ and\
  \bibinfo {author} {\bibfnamefont {J.}~\bibnamefont {Sampaio}},\ }\href@noop
  {} {\bibfield  {journal} {\bibinfo  {journal} {Nature nanotechnology}\
  }\textbf {\bibinfo {volume} {8}},\ \bibinfo {pages} {152} (\bibinfo {year}
  {2013})}\BibitemShut {NoStop}%
\bibitem [{\citenamefont {Fert}\ \emph {et~al.}(2017)\citenamefont {Fert},
  \citenamefont {Reyren},\ and\ \citenamefont {Cros}}]{fert2017magnetic}%
  \BibitemOpen
  \bibfield  {author} {\bibinfo {author} {\bibfnamefont {A.}~\bibnamefont
  {Fert}}, \bibinfo {author} {\bibfnamefont {N.}~\bibnamefont {Reyren}}, \ and\
  \bibinfo {author} {\bibfnamefont {V.}~\bibnamefont {Cros}},\ }\href@noop {}
  {\bibfield  {journal} {\bibinfo  {journal} {Nature Reviews Materials}\
  }\textbf {\bibinfo {volume} {2}},\ \bibinfo {pages} {1} (\bibinfo {year}
  {2017})}\BibitemShut {NoStop}%
\bibitem [{\citenamefont {Song}\ \emph {et~al.}(2020)\citenamefont {Song},
  \citenamefont {Jeong}, \citenamefont {Pan}, \citenamefont {Zhang},
  \citenamefont {Xia}, \citenamefont {Cha}, \citenamefont {Park}, \citenamefont
  {Kim}, \citenamefont {Finizio}, \citenamefont {Raabe} \emph
  {et~al.}}]{song2020skyrmion}%
  \BibitemOpen
  \bibfield  {author} {\bibinfo {author} {\bibfnamefont {K.~M.}\ \bibnamefont
  {Song}}, \bibinfo {author} {\bibfnamefont {J.-S.}\ \bibnamefont {Jeong}},
  \bibinfo {author} {\bibfnamefont {B.}~\bibnamefont {Pan}}, \bibinfo {author}
  {\bibfnamefont {X.}~\bibnamefont {Zhang}}, \bibinfo {author} {\bibfnamefont
  {J.}~\bibnamefont {Xia}}, \bibinfo {author} {\bibfnamefont {S.}~\bibnamefont
  {Cha}}, \bibinfo {author} {\bibfnamefont {T.-E.}\ \bibnamefont {Park}},
  \bibinfo {author} {\bibfnamefont {K.}~\bibnamefont {Kim}}, \bibinfo {author}
  {\bibfnamefont {S.}~\bibnamefont {Finizio}}, \bibinfo {author} {\bibfnamefont
  {J.}~\bibnamefont {Raabe}},  \emph {et~al.},\ }\href@noop {} {\bibfield
  {journal} {\bibinfo  {journal} {Nature Electronics}\ }\textbf {\bibinfo
  {volume} {3}},\ \bibinfo {pages} {148} (\bibinfo {year} {2020})}\BibitemShut
  {NoStop}%
\bibitem [{\citenamefont {Jonietz}\ \emph {et~al.}(2010)\citenamefont
  {Jonietz}, \citenamefont {M{\"u}hlbauer}, \citenamefont {Pfleiderer},
  \citenamefont {Neubauer}, \citenamefont {M{\"u}nzer}, \citenamefont {Bauer},
  \citenamefont {Adams}, \citenamefont {Georgii}, \citenamefont {B{\"o}ni},
  \citenamefont {Duine} \emph {et~al.}}]{jonietz2010spin}%
  \BibitemOpen
  \bibfield  {author} {\bibinfo {author} {\bibfnamefont {F.}~\bibnamefont
  {Jonietz}}, \bibinfo {author} {\bibfnamefont {S.}~\bibnamefont
  {M{\"u}hlbauer}}, \bibinfo {author} {\bibfnamefont {C.}~\bibnamefont
  {Pfleiderer}}, \bibinfo {author} {\bibfnamefont {A.}~\bibnamefont
  {Neubauer}}, \bibinfo {author} {\bibfnamefont {W.}~\bibnamefont
  {M{\"u}nzer}}, \bibinfo {author} {\bibfnamefont {A.}~\bibnamefont {Bauer}},
  \bibinfo {author} {\bibfnamefont {T.}~\bibnamefont {Adams}}, \bibinfo
  {author} {\bibfnamefont {R.}~\bibnamefont {Georgii}}, \bibinfo {author}
  {\bibfnamefont {P.}~\bibnamefont {B{\"o}ni}}, \bibinfo {author}
  {\bibfnamefont {R.~A.}\ \bibnamefont {Duine}},  \emph {et~al.},\ }\href@noop
  {} {\bibfield  {journal} {\bibinfo  {journal} {Science}\ }\textbf {\bibinfo
  {volume} {330}},\ \bibinfo {pages} {1648} (\bibinfo {year}
  {2010})}\BibitemShut {NoStop}%
\bibitem [{\citenamefont {Nagaosa}\ and\ \citenamefont
  {Tokura}(2013)}]{nagaosa2013topological}%
  \BibitemOpen
  \bibfield  {author} {\bibinfo {author} {\bibfnamefont {N.}~\bibnamefont
  {Nagaosa}}\ and\ \bibinfo {author} {\bibfnamefont {Y.}~\bibnamefont
  {Tokura}},\ }\href@noop {} {\bibfield  {journal} {\bibinfo  {journal} {Nature
  nanotechnology}\ }\textbf {\bibinfo {volume} {8}},\ \bibinfo {pages} {899}
  (\bibinfo {year} {2013})}\BibitemShut {NoStop}%
\bibitem [{\citenamefont {Jiang}\ \emph {et~al.}(2017)\citenamefont {Jiang},
  \citenamefont {Chen}, \citenamefont {Liu}, \citenamefont {Zang},
  \citenamefont {Te~Velthuis},\ and\ \citenamefont
  {Hoffmann}}]{jiang2017skyrmions}%
  \BibitemOpen
  \bibfield  {author} {\bibinfo {author} {\bibfnamefont {W.}~\bibnamefont
  {Jiang}}, \bibinfo {author} {\bibfnamefont {G.}~\bibnamefont {Chen}},
  \bibinfo {author} {\bibfnamefont {K.}~\bibnamefont {Liu}}, \bibinfo {author}
  {\bibfnamefont {J.}~\bibnamefont {Zang}}, \bibinfo {author} {\bibfnamefont
  {S.~G.}\ \bibnamefont {Te~Velthuis}}, \ and\ \bibinfo {author} {\bibfnamefont
  {A.}~\bibnamefont {Hoffmann}},\ }\href@noop {} {\bibfield  {journal}
  {\bibinfo  {journal} {Physics Reports}\ }\textbf {\bibinfo {volume} {704}},\
  \bibinfo {pages} {1} (\bibinfo {year} {2017})}\BibitemShut {NoStop}%
\bibitem [{\citenamefont {Jiang}\ \emph {et~al.}(2020)\citenamefont {Jiang},
  \citenamefont {Xiao}, \citenamefont {Wang}, \citenamefont {Shin},
  \citenamefont {Andreoli}, \citenamefont {Zhang}, \citenamefont {Xiao},
  \citenamefont {Zhao}, \citenamefont {Kayyalha}, \citenamefont {Zhang} \emph
  {et~al.}}]{jiang2020concurrence}%
  \BibitemOpen
  \bibfield  {author} {\bibinfo {author} {\bibfnamefont {J.}~\bibnamefont
  {Jiang}}, \bibinfo {author} {\bibfnamefont {D.}~\bibnamefont {Xiao}},
  \bibinfo {author} {\bibfnamefont {F.}~\bibnamefont {Wang}}, \bibinfo {author}
  {\bibfnamefont {J.-H.}\ \bibnamefont {Shin}}, \bibinfo {author}
  {\bibfnamefont {D.}~\bibnamefont {Andreoli}}, \bibinfo {author}
  {\bibfnamefont {J.}~\bibnamefont {Zhang}}, \bibinfo {author} {\bibfnamefont
  {R.}~\bibnamefont {Xiao}}, \bibinfo {author} {\bibfnamefont {Y.-F.}\
  \bibnamefont {Zhao}}, \bibinfo {author} {\bibfnamefont {M.}~\bibnamefont
  {Kayyalha}}, \bibinfo {author} {\bibfnamefont {L.}~\bibnamefont {Zhang}},
  \emph {et~al.},\ }\href@noop {} {\bibfield  {journal} {\bibinfo  {journal}
  {Nature Materials}\ }\textbf {\bibinfo {volume} {19}},\ \bibinfo {pages}
  {732} (\bibinfo {year} {2020})}\BibitemShut {NoStop}%
\bibitem [{\citenamefont {Li}\ \emph {et~al.}(2022)\citenamefont {Li},
  \citenamefont {Xu}, \citenamefont {Wang}, \citenamefont {Wang}, \citenamefont
  {Yang}, \citenamefont {Lin}, \citenamefont {Duan},\ and\ \citenamefont
  {Huang}}]{li2022interplay}%
  \BibitemOpen
  \bibfield  {author} {\bibinfo {author} {\bibfnamefont {Y.}~\bibnamefont
  {Li}}, \bibinfo {author} {\bibfnamefont {S.}~\bibnamefont {Xu}}, \bibinfo
  {author} {\bibfnamefont {J.}~\bibnamefont {Wang}}, \bibinfo {author}
  {\bibfnamefont {C.}~\bibnamefont {Wang}}, \bibinfo {author} {\bibfnamefont
  {B.}~\bibnamefont {Yang}}, \bibinfo {author} {\bibfnamefont {H.}~\bibnamefont
  {Lin}}, \bibinfo {author} {\bibfnamefont {W.}~\bibnamefont {Duan}}, \ and\
  \bibinfo {author} {\bibfnamefont {B.}~\bibnamefont {Huang}},\ }\href@noop {}
  {\bibfield  {journal} {\bibinfo  {journal} {Proceedings of the National
  Academy of Sciences}\ }\textbf {\bibinfo {volume} {119}},\ \bibinfo {pages}
  {e2122952119} (\bibinfo {year} {2022})}\BibitemShut {NoStop}%
\bibitem [{\citenamefont {Li}\ \emph {et~al.}(2020)\citenamefont {Li},
  \citenamefont {Ding}, \citenamefont {Zhang}, \citenamefont {Kally},
  \citenamefont {Pillsbury}, \citenamefont {Heinonen}, \citenamefont {Rimal},
  \citenamefont {Bi}, \citenamefont {DeMann}, \citenamefont {Field} \emph
  {et~al.}}]{li2020topological}%
  \BibitemOpen
  \bibfield  {author} {\bibinfo {author} {\bibfnamefont {P.}~\bibnamefont
  {Li}}, \bibinfo {author} {\bibfnamefont {J.}~\bibnamefont {Ding}}, \bibinfo
  {author} {\bibfnamefont {S.~S.-L.}\ \bibnamefont {Zhang}}, \bibinfo {author}
  {\bibfnamefont {J.}~\bibnamefont {Kally}}, \bibinfo {author} {\bibfnamefont
  {T.}~\bibnamefont {Pillsbury}}, \bibinfo {author} {\bibfnamefont {O.~G.}\
  \bibnamefont {Heinonen}}, \bibinfo {author} {\bibfnamefont {G.}~\bibnamefont
  {Rimal}}, \bibinfo {author} {\bibfnamefont {C.}~\bibnamefont {Bi}}, \bibinfo
  {author} {\bibfnamefont {A.}~\bibnamefont {DeMann}}, \bibinfo {author}
  {\bibfnamefont {S.~B.}\ \bibnamefont {Field}},  \emph {et~al.},\ }\href@noop
  {} {\bibfield  {journal} {\bibinfo  {journal} {Nano letters}\ }\textbf
  {\bibinfo {volume} {21}},\ \bibinfo {pages} {84} (\bibinfo {year}
  {2020})}\BibitemShut {NoStop}%
\bibitem [{\citenamefont {Zou}\ \emph {et~al.}(2022)\citenamefont {Zou},
  \citenamefont {Guo}, \citenamefont {Wong}, \citenamefont {Huang},
  \citenamefont {Chia}, \citenamefont {Chen}, \citenamefont {Wang},
  \citenamefont {Lin}, \citenamefont {Young}, \citenamefont {Lin} \emph
  {et~al.}}]{zou2022enormous}%
  \BibitemOpen
  \bibfield  {author} {\bibinfo {author} {\bibfnamefont {W.-J.}\ \bibnamefont
  {Zou}}, \bibinfo {author} {\bibfnamefont {M.-X.}\ \bibnamefont {Guo}},
  \bibinfo {author} {\bibfnamefont {J.-F.}\ \bibnamefont {Wong}}, \bibinfo
  {author} {\bibfnamefont {Z.-P.}\ \bibnamefont {Huang}}, \bibinfo {author}
  {\bibfnamefont {J.-M.}\ \bibnamefont {Chia}}, \bibinfo {author}
  {\bibfnamefont {W.-N.}\ \bibnamefont {Chen}}, \bibinfo {author}
  {\bibfnamefont {S.-X.}\ \bibnamefont {Wang}}, \bibinfo {author}
  {\bibfnamefont {K.-Y.}\ \bibnamefont {Lin}}, \bibinfo {author} {\bibfnamefont
  {L.~B.}\ \bibnamefont {Young}}, \bibinfo {author} {\bibfnamefont {Y.-H.~G.}\
  \bibnamefont {Lin}},  \emph {et~al.},\ }\href@noop {} {\bibfield  {journal}
  {\bibinfo  {journal} {ACS nano}\ }\textbf {\bibinfo {volume} {16}},\ \bibinfo
  {pages} {2369} (\bibinfo {year} {2022})}\BibitemShut {NoStop}%
\bibitem [{\citenamefont {Xiao}\ \emph {et~al.}(2021)\citenamefont {Xiao},
  \citenamefont {Xiao}, \citenamefont {Jiang}, \citenamefont {Shin},
  \citenamefont {Wang}, \citenamefont {Zhao}, \citenamefont {Zhang},
  \citenamefont {Richardella}, \citenamefont {Wang}, \citenamefont {Kayyalha}
  \emph {et~al.}}]{xiao2021mapping}%
  \BibitemOpen
  \bibfield  {author} {\bibinfo {author} {\bibfnamefont {R.}~\bibnamefont
  {Xiao}}, \bibinfo {author} {\bibfnamefont {D.}~\bibnamefont {Xiao}}, \bibinfo
  {author} {\bibfnamefont {J.}~\bibnamefont {Jiang}}, \bibinfo {author}
  {\bibfnamefont {J.-H.}\ \bibnamefont {Shin}}, \bibinfo {author}
  {\bibfnamefont {F.}~\bibnamefont {Wang}}, \bibinfo {author} {\bibfnamefont
  {Y.-F.}\ \bibnamefont {Zhao}}, \bibinfo {author} {\bibfnamefont {R.-X.}\
  \bibnamefont {Zhang}}, \bibinfo {author} {\bibfnamefont {A.}~\bibnamefont
  {Richardella}}, \bibinfo {author} {\bibfnamefont {K.}~\bibnamefont {Wang}},
  \bibinfo {author} {\bibfnamefont {M.}~\bibnamefont {Kayyalha}},  \emph
  {et~al.},\ }\href@noop {} {\bibfield  {journal} {\bibinfo  {journal}
  {Physical Review Research}\ }\textbf {\bibinfo {volume} {3}},\ \bibinfo
  {pages} {L032004} (\bibinfo {year} {2021})}\BibitemShut {NoStop}%
\bibitem [{\citenamefont {Kawasaki}\ \emph {et~al.}(2016)\citenamefont
  {Kawasaki}, \citenamefont {Kaneko}, \citenamefont {Nakamura}, \citenamefont
  {Aso}, \citenamefont {Hedo}, \citenamefont {Nakama}, \citenamefont {Ohhara},
  \citenamefont {Kiyanagi}, \citenamefont {Oikawa}, \citenamefont {Tamura}
  \emph {et~al.}}]{kawasaki2016magnetic}%
  \BibitemOpen
  \bibfield  {author} {\bibinfo {author} {\bibfnamefont {T.}~\bibnamefont
  {Kawasaki}}, \bibinfo {author} {\bibfnamefont {K.}~\bibnamefont {Kaneko}},
  \bibinfo {author} {\bibfnamefont {A.}~\bibnamefont {Nakamura}}, \bibinfo
  {author} {\bibfnamefont {N.}~\bibnamefont {Aso}}, \bibinfo {author}
  {\bibfnamefont {M.}~\bibnamefont {Hedo}}, \bibinfo {author} {\bibfnamefont
  {T.}~\bibnamefont {Nakama}}, \bibinfo {author} {\bibfnamefont
  {T.}~\bibnamefont {Ohhara}}, \bibinfo {author} {\bibfnamefont
  {R.}~\bibnamefont {Kiyanagi}}, \bibinfo {author} {\bibfnamefont
  {K.}~\bibnamefont {Oikawa}}, \bibinfo {author} {\bibfnamefont
  {I.}~\bibnamefont {Tamura}},  \emph {et~al.},\ }\href@noop {} {\bibfield
  {journal} {\bibinfo  {journal} {Journal of the Physical Society of Japan}\
  }\textbf {\bibinfo {volume} {85}},\ \bibinfo {pages} {114711} (\bibinfo
  {year} {2016})}\BibitemShut {NoStop}%
\bibitem [{\citenamefont {Lei}\ \emph {et~al.}(2022)\citenamefont {Lei},
  \citenamefont {Allen}, \citenamefont {Huang}, \citenamefont {Moya},
  \citenamefont {Casas}, \citenamefont {Zhang}, \citenamefont {Hashimoto},
  \citenamefont {Lu}, \citenamefont {Denlinger}, \citenamefont {Balicas},
  \citenamefont {Yi}, \citenamefont {Sun},\ and\ \citenamefont
  {Morosan}}]{Kevin2021}%
  \BibitemOpen
  \bibfield  {author} {\bibinfo {author} {\bibfnamefont {S.}~\bibnamefont
  {Lei}}, \bibinfo {author} {\bibfnamefont {K.}~\bibnamefont {Allen}}, \bibinfo
  {author} {\bibfnamefont {J.}~\bibnamefont {Huang}}, \bibinfo {author}
  {\bibfnamefont {J.~M.}\ \bibnamefont {Moya}}, \bibinfo {author}
  {\bibfnamefont {B.}~\bibnamefont {Casas}}, \bibinfo {author} {\bibfnamefont
  {Y.}~\bibnamefont {Zhang}}, \bibinfo {author} {\bibfnamefont
  {M.}~\bibnamefont {Hashimoto}}, \bibinfo {author} {\bibfnamefont
  {D.}~\bibnamefont {Lu}}, \bibinfo {author} {\bibfnamefont {J.}~\bibnamefont
  {Denlinger}}, \bibinfo {author} {\bibfnamefont {L.}~\bibnamefont {Balicas}},
  \bibinfo {author} {\bibfnamefont {M.}~\bibnamefont {Yi}}, \bibinfo {author}
  {\bibfnamefont {Y.}~\bibnamefont {Sun}}, \ and\ \bibinfo {author}
  {\bibfnamefont {E.}~\bibnamefont {Morosan}},\ }\href {\doibase
  10.48550/ARXIV.2208.06407} {\enquote {\bibinfo {title} {Weyl nodal ring
  states and landau quantization with very large magnetoresistance in
  square-net magnet euga$_4$},}\ } (\bibinfo {year} {2022})\BibitemShut
  {NoStop}%
\bibitem [{\citenamefont {Zhang}\ \emph {et~al.}(2021)\citenamefont {Zhang},
  \citenamefont {Zhu}, \citenamefont {Xu}, \citenamefont {Gawryluk},
  \citenamefont {Xie}, \citenamefont {Ju}, \citenamefont {Shi}, \citenamefont
  {Shiroka}, \citenamefont {Zhan}, \citenamefont {Pomjakushina} \emph
  {et~al.}}]{zhang2021giant}%
  \BibitemOpen
  \bibfield  {author} {\bibinfo {author} {\bibfnamefont {H.}~\bibnamefont
  {Zhang}}, \bibinfo {author} {\bibfnamefont {X.}~\bibnamefont {Zhu}}, \bibinfo
  {author} {\bibfnamefont {Y.}~\bibnamefont {Xu}}, \bibinfo {author}
  {\bibfnamefont {D.}~\bibnamefont {Gawryluk}}, \bibinfo {author}
  {\bibfnamefont {W.}~\bibnamefont {Xie}}, \bibinfo {author} {\bibfnamefont
  {S.}~\bibnamefont {Ju}}, \bibinfo {author} {\bibfnamefont {M.}~\bibnamefont
  {Shi}}, \bibinfo {author} {\bibfnamefont {T.}~\bibnamefont {Shiroka}},
  \bibinfo {author} {\bibfnamefont {Q.}~\bibnamefont {Zhan}}, \bibinfo {author}
  {\bibfnamefont {E.}~\bibnamefont {Pomjakushina}},  \emph {et~al.},\
  }\href@noop {} {\bibfield  {journal} {\bibinfo  {journal} {Journal of
  Physics: Condensed Matter}\ }\textbf {\bibinfo {volume} {34}},\ \bibinfo
  {pages} {034005} (\bibinfo {year} {2021})}\BibitemShut {NoStop}%
\bibitem [{\citenamefont {Stavinoha}\ \emph {et~al.}(2018)\citenamefont
  {Stavinoha}, \citenamefont {Cooley}, \citenamefont {Minasian}, \citenamefont
  {McQueen}, \citenamefont {Kauzlarich}, \citenamefont {Huang},\ and\
  \citenamefont {Morosan}}]{stavinoha2018charge}%
  \BibitemOpen
  \bibfield  {author} {\bibinfo {author} {\bibfnamefont {M.}~\bibnamefont
  {Stavinoha}}, \bibinfo {author} {\bibfnamefont {J.~A.}\ \bibnamefont
  {Cooley}}, \bibinfo {author} {\bibfnamefont {S.~G.}\ \bibnamefont
  {Minasian}}, \bibinfo {author} {\bibfnamefont {T.~M.}\ \bibnamefont
  {McQueen}}, \bibinfo {author} {\bibfnamefont {S.~M.}\ \bibnamefont
  {Kauzlarich}}, \bibinfo {author} {\bibfnamefont {C.-L.}\ \bibnamefont
  {Huang}}, \ and\ \bibinfo {author} {\bibfnamefont {E.}~\bibnamefont
  {Morosan}},\ }\href@noop {} {\bibfield  {journal} {\bibinfo  {journal}
  {Physical Review B}\ }\textbf {\bibinfo {volume} {97}},\ \bibinfo {pages}
  {195146} (\bibinfo {year} {2018})}\BibitemShut {NoStop}%
\bibitem [{\citenamefont {Takagi}\ \emph {et~al.}(2022)\citenamefont {Takagi},
  \citenamefont {Matsuyama}, \citenamefont {Ukleev}, \citenamefont {Yu},
  \citenamefont {White}, \citenamefont {Francoual}, \citenamefont {Mardegan},
  \citenamefont {Hayami}, \citenamefont {Saito}, \citenamefont {Kaneko} \emph
  {et~al.}}]{takagi2022square}%
  \BibitemOpen
  \bibfield  {author} {\bibinfo {author} {\bibfnamefont {R.}~\bibnamefont
  {Takagi}}, \bibinfo {author} {\bibfnamefont {N.}~\bibnamefont {Matsuyama}},
  \bibinfo {author} {\bibfnamefont {V.}~\bibnamefont {Ukleev}}, \bibinfo
  {author} {\bibfnamefont {L.}~\bibnamefont {Yu}}, \bibinfo {author}
  {\bibfnamefont {J.~S.}\ \bibnamefont {White}}, \bibinfo {author}
  {\bibfnamefont {S.}~\bibnamefont {Francoual}}, \bibinfo {author}
  {\bibfnamefont {J.~R.}\ \bibnamefont {Mardegan}}, \bibinfo {author}
  {\bibfnamefont {S.}~\bibnamefont {Hayami}}, \bibinfo {author} {\bibfnamefont
  {H.}~\bibnamefont {Saito}}, \bibinfo {author} {\bibfnamefont
  {K.}~\bibnamefont {Kaneko}},  \emph {et~al.},\ }\href@noop {} {\bibfield
  {journal} {\bibinfo  {journal} {Nature Communications}\ }\textbf {\bibinfo
  {volume} {13}},\ \bibinfo {pages} {1} (\bibinfo {year} {2022})}\BibitemShut
  {NoStop}%
\bibitem [{\citenamefont {Shang}\ \emph {et~al.}(2021)\citenamefont {Shang},
  \citenamefont {Xu}, \citenamefont {Gawryluk}, \citenamefont {Ma},
  \citenamefont {Shiroka}, \citenamefont {Shi},\ and\ \citenamefont
  {Pomjakushina}}]{shang2021anomalous}%
  \BibitemOpen
  \bibfield  {author} {\bibinfo {author} {\bibfnamefont {T.}~\bibnamefont
  {Shang}}, \bibinfo {author} {\bibfnamefont {Y.}~\bibnamefont {Xu}}, \bibinfo
  {author} {\bibfnamefont {D.}~\bibnamefont {Gawryluk}}, \bibinfo {author}
  {\bibfnamefont {J.}~\bibnamefont {Ma}}, \bibinfo {author} {\bibfnamefont
  {T.}~\bibnamefont {Shiroka}}, \bibinfo {author} {\bibfnamefont
  {M.}~\bibnamefont {Shi}}, \ and\ \bibinfo {author} {\bibfnamefont
  {E.}~\bibnamefont {Pomjakushina}},\ }\href@noop {} {\bibfield  {journal}
  {\bibinfo  {journal} {Physical Review B}\ }\textbf {\bibinfo {volume}
  {103}},\ \bibinfo {pages} {L020405} (\bibinfo {year} {2021})}\BibitemShut
  {NoStop}%
\bibitem [{\citenamefont {Meier}\ \emph {et~al.}(2022)\citenamefont {Meier},
  \citenamefont {Torres}, \citenamefont {Hermann}, \citenamefont {Zhao},
  \citenamefont {Lavina}, \citenamefont {Sales},\ and\ \citenamefont
  {May}}]{meier2022thermodynamic}%
  \BibitemOpen
  \bibfield  {author} {\bibinfo {author} {\bibfnamefont {W.~R.}\ \bibnamefont
  {Meier}}, \bibinfo {author} {\bibfnamefont {J.~R.}\ \bibnamefont {Torres}},
  \bibinfo {author} {\bibfnamefont {R.~P.}\ \bibnamefont {Hermann}}, \bibinfo
  {author} {\bibfnamefont {J.}~\bibnamefont {Zhao}}, \bibinfo {author}
  {\bibfnamefont {B.}~\bibnamefont {Lavina}}, \bibinfo {author} {\bibfnamefont
  {B.~C.}\ \bibnamefont {Sales}}, \ and\ \bibinfo {author} {\bibfnamefont
  {A.~F.}\ \bibnamefont {May}},\ }\href@noop {} {\bibfield  {journal} {\bibinfo
   {journal} {arXiv preprint arXiv:2204.02319}\ } (\bibinfo {year}
  {2022})}\BibitemShut {NoStop}%
\bibitem [{\citenamefont {Kaneko}\ \emph {et~al.}(2021)\citenamefont {Kaneko},
  \citenamefont {Kawasaki}, \citenamefont {Nakamura}, \citenamefont {Munakata},
  \citenamefont {Nakao}, \citenamefont {Hanashima}, \citenamefont {Kiyanagi},
  \citenamefont {Ohhara}, \citenamefont {Hedo}, \citenamefont {Nakama} \emph
  {et~al.}}]{kaneko2021charge}%
  \BibitemOpen
  \bibfield  {author} {\bibinfo {author} {\bibfnamefont {K.}~\bibnamefont
  {Kaneko}}, \bibinfo {author} {\bibfnamefont {T.}~\bibnamefont {Kawasaki}},
  \bibinfo {author} {\bibfnamefont {A.}~\bibnamefont {Nakamura}}, \bibinfo
  {author} {\bibfnamefont {K.}~\bibnamefont {Munakata}}, \bibinfo {author}
  {\bibfnamefont {A.}~\bibnamefont {Nakao}}, \bibinfo {author} {\bibfnamefont
  {T.}~\bibnamefont {Hanashima}}, \bibinfo {author} {\bibfnamefont
  {R.}~\bibnamefont {Kiyanagi}}, \bibinfo {author} {\bibfnamefont
  {T.}~\bibnamefont {Ohhara}}, \bibinfo {author} {\bibfnamefont
  {M.}~\bibnamefont {Hedo}}, \bibinfo {author} {\bibfnamefont {T.}~\bibnamefont
  {Nakama}},  \emph {et~al.},\ }\href@noop {} {\bibfield  {journal} {\bibinfo
  {journal} {Journal of the Physical Society of Japan}\ }\textbf {\bibinfo
  {volume} {90}},\ \bibinfo {pages} {064704} (\bibinfo {year}
  {2021})}\BibitemShut {NoStop}%
\bibitem [{\citenamefont {Shimomura}\ \emph {et~al.}(2019)\citenamefont
  {Shimomura}, \citenamefont {Murao}, \citenamefont {Tsutsui}, \citenamefont
  {Nakao}, \citenamefont {Nakamura}, \citenamefont {Hedo}, \citenamefont
  {Nakama},\ and\ \citenamefont {{\=O}nuki}}]{shimomura2019lattice}%
  \BibitemOpen
  \bibfield  {author} {\bibinfo {author} {\bibfnamefont {S.}~\bibnamefont
  {Shimomura}}, \bibinfo {author} {\bibfnamefont {H.}~\bibnamefont {Murao}},
  \bibinfo {author} {\bibfnamefont {S.}~\bibnamefont {Tsutsui}}, \bibinfo
  {author} {\bibfnamefont {H.}~\bibnamefont {Nakao}}, \bibinfo {author}
  {\bibfnamefont {A.}~\bibnamefont {Nakamura}}, \bibinfo {author}
  {\bibfnamefont {M.}~\bibnamefont {Hedo}}, \bibinfo {author} {\bibfnamefont
  {T.}~\bibnamefont {Nakama}}, \ and\ \bibinfo {author} {\bibfnamefont
  {Y.}~\bibnamefont {{\=O}nuki}},\ }\href@noop {} {\bibfield  {journal}
  {\bibinfo  {journal} {Journal of the Physical Society of Japan}\ }\textbf
  {\bibinfo {volume} {88}},\ \bibinfo {pages} {014602} (\bibinfo {year}
  {2019})}\BibitemShut {NoStop}%
\bibitem [{\citenamefont {Moya}\ \emph {et~al.}(2022)\citenamefont {Moya},
  \citenamefont {Lei}, \citenamefont {Clements}, \citenamefont {Kengle},
  \citenamefont {Sun}, \citenamefont {Allen}, \citenamefont {Li}, \citenamefont
  {Peng}, \citenamefont {Husain}, \citenamefont {Mitrano} \emph
  {et~al.}}]{moya2022incommensurate}%
  \BibitemOpen
  \bibfield  {author} {\bibinfo {author} {\bibfnamefont {J.~M.}\ \bibnamefont
  {Moya}}, \bibinfo {author} {\bibfnamefont {S.}~\bibnamefont {Lei}}, \bibinfo
  {author} {\bibfnamefont {E.~M.}\ \bibnamefont {Clements}}, \bibinfo {author}
  {\bibfnamefont {C.~S.}\ \bibnamefont {Kengle}}, \bibinfo {author}
  {\bibfnamefont {S.}~\bibnamefont {Sun}}, \bibinfo {author} {\bibfnamefont
  {K.}~\bibnamefont {Allen}}, \bibinfo {author} {\bibfnamefont
  {Q.}~\bibnamefont {Li}}, \bibinfo {author} {\bibfnamefont {Y.}~\bibnamefont
  {Peng}}, \bibinfo {author} {\bibfnamefont {A.~A.}\ \bibnamefont {Husain}},
  \bibinfo {author} {\bibfnamefont {M.}~\bibnamefont {Mitrano}},  \emph
  {et~al.},\ }\href@noop {} {\bibfield  {journal} {\bibinfo  {journal}
  {Physical Review Materials}\ }\textbf {\bibinfo {volume} {6}},\ \bibinfo
  {pages} {074201} (\bibinfo {year} {2022})}\BibitemShut {NoStop}%
\bibitem [{\citenamefont
  {Rodr{\'\i}guez-Carvajal}(1993)}]{rodriguez1993recent}%
  \BibitemOpen
  \bibfield  {author} {\bibinfo {author} {\bibfnamefont {J.}~\bibnamefont
  {Rodr{\'\i}guez-Carvajal}},\ }\href@noop {} {\bibfield  {journal} {\bibinfo
  {journal} {Physica B: Condensed Matter}\ }\textbf {\bibinfo {volume} {192}},\
  \bibinfo {pages} {55} (\bibinfo {year} {1993})}\BibitemShut {NoStop}%
\bibitem [{\citenamefont {Kresse}\ and\ \citenamefont
  {Furthm{\"u}ller}(1996)}]{kresse1996efficient}%
  \BibitemOpen
  \bibfield  {author} {\bibinfo {author} {\bibfnamefont {G.}~\bibnamefont
  {Kresse}}\ and\ \bibinfo {author} {\bibfnamefont {J.}~\bibnamefont
  {Furthm{\"u}ller}},\ }\href@noop {} {\bibfield  {journal} {\bibinfo
  {journal} {Physical Review B}\ }\textbf {\bibinfo {volume} {54}},\ \bibinfo
  {pages} {11169} (\bibinfo {year} {1996})}\BibitemShut {NoStop}%
\bibitem [{\citenamefont {Perdew}\ \emph {et~al.}(1996)\citenamefont {Perdew},
  \citenamefont {Burke},\ and\ \citenamefont
  {Ernzerhof}}]{perdew1996generalized}%
  \BibitemOpen
  \bibfield  {author} {\bibinfo {author} {\bibfnamefont {J.~P.}\ \bibnamefont
  {Perdew}}, \bibinfo {author} {\bibfnamefont {K.}~\bibnamefont {Burke}}, \
  and\ \bibinfo {author} {\bibfnamefont {M.}~\bibnamefont {Ernzerhof}},\
  }\href@noop {} {\bibfield  {journal} {\bibinfo  {journal} {Physical review
  letters}\ }\textbf {\bibinfo {volume} {77}},\ \bibinfo {pages} {3865}
  (\bibinfo {year} {1996})}\BibitemShut {NoStop}%
\bibitem [{\citenamefont {Mostofi}\ \emph {et~al.}(2008)\citenamefont
  {Mostofi}, \citenamefont {Yates}, \citenamefont {Lee}, \citenamefont {Souza},
  \citenamefont {Vanderbilt},\ and\ \citenamefont
  {Marzari}}]{mostofi2008wannier90}%
  \BibitemOpen
  \bibfield  {author} {\bibinfo {author} {\bibfnamefont {A.~A.}\ \bibnamefont
  {Mostofi}}, \bibinfo {author} {\bibfnamefont {J.~R.}\ \bibnamefont {Yates}},
  \bibinfo {author} {\bibfnamefont {Y.-S.}\ \bibnamefont {Lee}}, \bibinfo
  {author} {\bibfnamefont {I.}~\bibnamefont {Souza}}, \bibinfo {author}
  {\bibfnamefont {D.}~\bibnamefont {Vanderbilt}}, \ and\ \bibinfo {author}
  {\bibfnamefont {N.}~\bibnamefont {Marzari}},\ }\href@noop {} {\bibfield
  {journal} {\bibinfo  {journal} {Computer physics communications}\ }\textbf
  {\bibinfo {volume} {178}},\ \bibinfo {pages} {685} (\bibinfo {year}
  {2008})}\BibitemShut {NoStop}%
\bibitem [{\citenamefont {Sancho}\ \emph {et~al.}(1984)\citenamefont {Sancho},
  \citenamefont {Sancho},\ and\ \citenamefont {Rubio}}]{sancho1984quick}%
  \BibitemOpen
  \bibfield  {author} {\bibinfo {author} {\bibfnamefont {M.~L.}\ \bibnamefont
  {Sancho}}, \bibinfo {author} {\bibfnamefont {J.~L.}\ \bibnamefont {Sancho}},
  \ and\ \bibinfo {author} {\bibfnamefont {J.}~\bibnamefont {Rubio}},\
  }\href@noop {} {\bibfield  {journal} {\bibinfo  {journal} {Journal of Physics
  F: Metal Physics}\ }\textbf {\bibinfo {volume} {14}},\ \bibinfo {pages}
  {1205} (\bibinfo {year} {1984})}\BibitemShut {NoStop}%
\bibitem [{\citenamefont {Sancho}\ \emph {et~al.}(1985)\citenamefont {Sancho},
  \citenamefont {Sancho}, \citenamefont {Sancho},\ and\ \citenamefont
  {Rubio}}]{sancho1985highly}%
  \BibitemOpen
  \bibfield  {author} {\bibinfo {author} {\bibfnamefont {M.~L.}\ \bibnamefont
  {Sancho}}, \bibinfo {author} {\bibfnamefont {J.~L.}\ \bibnamefont {Sancho}},
  \bibinfo {author} {\bibfnamefont {J.~L.}\ \bibnamefont {Sancho}}, \ and\
  \bibinfo {author} {\bibfnamefont {J.}~\bibnamefont {Rubio}},\ }\href@noop {}
  {\bibfield  {journal} {\bibinfo  {journal} {Journal of Physics F: Metal
  Physics}\ }\textbf {\bibinfo {volume} {15}},\ \bibinfo {pages} {851}
  (\bibinfo {year} {1985})}\BibitemShut {NoStop}%
\bibitem [{\citenamefont {Ye}\ \emph {et~al.}(2018)\citenamefont {Ye},
  \citenamefont {Kang}, \citenamefont {Liu}, \citenamefont {Von~Cube},
  \citenamefont {Wicker}, \citenamefont {Suzuki}, \citenamefont {Jozwiak},
  \citenamefont {Bostwick}, \citenamefont {Rotenberg}, \citenamefont {Bell}
  \emph {et~al.}}]{ye2018massive}%
  \BibitemOpen
  \bibfield  {author} {\bibinfo {author} {\bibfnamefont {L.}~\bibnamefont
  {Ye}}, \bibinfo {author} {\bibfnamefont {M.}~\bibnamefont {Kang}}, \bibinfo
  {author} {\bibfnamefont {J.}~\bibnamefont {Liu}}, \bibinfo {author}
  {\bibfnamefont {F.}~\bibnamefont {Von~Cube}}, \bibinfo {author}
  {\bibfnamefont {C.~R.}\ \bibnamefont {Wicker}}, \bibinfo {author}
  {\bibfnamefont {T.}~\bibnamefont {Suzuki}}, \bibinfo {author} {\bibfnamefont
  {C.}~\bibnamefont {Jozwiak}}, \bibinfo {author} {\bibfnamefont
  {A.}~\bibnamefont {Bostwick}}, \bibinfo {author} {\bibfnamefont
  {E.}~\bibnamefont {Rotenberg}}, \bibinfo {author} {\bibfnamefont {D.~C.}\
  \bibnamefont {Bell}},  \emph {et~al.},\ }\href@noop {} {\bibfield  {journal}
  {\bibinfo  {journal} {Nature}\ }\textbf {\bibinfo {volume} {555}},\ \bibinfo
  {pages} {638} (\bibinfo {year} {2018})}\BibitemShut {NoStop}%
\bibitem [{\citenamefont {Suzuki}\ \emph {et~al.}(2016)\citenamefont {Suzuki},
  \citenamefont {Chisnell}, \citenamefont {Devarakonda}, \citenamefont {Liu},
  \citenamefont {Feng}, \citenamefont {Xiao}, \citenamefont {Lynn},\ and\
  \citenamefont {Checkelsky}}]{suzuki2016large}%
  \BibitemOpen
  \bibfield  {author} {\bibinfo {author} {\bibfnamefont {T.}~\bibnamefont
  {Suzuki}}, \bibinfo {author} {\bibfnamefont {R.}~\bibnamefont {Chisnell}},
  \bibinfo {author} {\bibfnamefont {A.}~\bibnamefont {Devarakonda}}, \bibinfo
  {author} {\bibfnamefont {Y.-T.}\ \bibnamefont {Liu}}, \bibinfo {author}
  {\bibfnamefont {W.}~\bibnamefont {Feng}}, \bibinfo {author} {\bibfnamefont
  {D.}~\bibnamefont {Xiao}}, \bibinfo {author} {\bibfnamefont {J.~W.}\
  \bibnamefont {Lynn}}, \ and\ \bibinfo {author} {\bibfnamefont
  {J.}~\bibnamefont {Checkelsky}},\ }\href@noop {} {\bibfield  {journal}
  {\bibinfo  {journal} {Nature Physics}\ }\textbf {\bibinfo {volume} {12}},\
  \bibinfo {pages} {1119} (\bibinfo {year} {2016})}\BibitemShut {NoStop}%
\bibitem [{\citenamefont {Liu}\ \emph {et~al.}(2018)\citenamefont {Liu},
  \citenamefont {Sun}, \citenamefont {Kumar}, \citenamefont {Muechler},
  \citenamefont {Sun}, \citenamefont {Jiao}, \citenamefont {Yang},
  \citenamefont {Liu}, \citenamefont {Liang}, \citenamefont {Xu} \emph
  {et~al.}}]{liu2018giant}%
  \BibitemOpen
  \bibfield  {author} {\bibinfo {author} {\bibfnamefont {E.}~\bibnamefont
  {Liu}}, \bibinfo {author} {\bibfnamefont {Y.}~\bibnamefont {Sun}}, \bibinfo
  {author} {\bibfnamefont {N.}~\bibnamefont {Kumar}}, \bibinfo {author}
  {\bibfnamefont {L.}~\bibnamefont {Muechler}}, \bibinfo {author}
  {\bibfnamefont {A.}~\bibnamefont {Sun}}, \bibinfo {author} {\bibfnamefont
  {L.}~\bibnamefont {Jiao}}, \bibinfo {author} {\bibfnamefont {S.-Y.}\
  \bibnamefont {Yang}}, \bibinfo {author} {\bibfnamefont {D.}~\bibnamefont
  {Liu}}, \bibinfo {author} {\bibfnamefont {A.}~\bibnamefont {Liang}}, \bibinfo
  {author} {\bibfnamefont {Q.}~\bibnamefont {Xu}},  \emph {et~al.},\
  }\href@noop {} {\bibfield  {journal} {\bibinfo  {journal} {Nature physics}\
  }\textbf {\bibinfo {volume} {14}},\ \bibinfo {pages} {1125} (\bibinfo {year}
  {2018})}\BibitemShut {NoStop}%
\bibitem [{\citenamefont {Yang}\ \emph {et~al.}(2020)\citenamefont {Yang},
  \citenamefont {Wang}, \citenamefont {Ortiz}, \citenamefont {Liu},
  \citenamefont {Gayles}, \citenamefont {Derunova}, \citenamefont
  {Gonzalez-Hernandez}, \citenamefont {{\v{S}}mejkal}, \citenamefont {Chen},
  \citenamefont {Parkin} \emph {et~al.}}]{yang2020giant}%
  \BibitemOpen
  \bibfield  {author} {\bibinfo {author} {\bibfnamefont {S.-Y.}\ \bibnamefont
  {Yang}}, \bibinfo {author} {\bibfnamefont {Y.}~\bibnamefont {Wang}}, \bibinfo
  {author} {\bibfnamefont {B.~R.}\ \bibnamefont {Ortiz}}, \bibinfo {author}
  {\bibfnamefont {D.}~\bibnamefont {Liu}}, \bibinfo {author} {\bibfnamefont
  {J.}~\bibnamefont {Gayles}}, \bibinfo {author} {\bibfnamefont
  {E.}~\bibnamefont {Derunova}}, \bibinfo {author} {\bibfnamefont
  {R.}~\bibnamefont {Gonzalez-Hernandez}}, \bibinfo {author} {\bibfnamefont
  {L.}~\bibnamefont {{\v{S}}mejkal}}, \bibinfo {author} {\bibfnamefont
  {Y.}~\bibnamefont {Chen}}, \bibinfo {author} {\bibfnamefont {S.~S.}\
  \bibnamefont {Parkin}},  \emph {et~al.},\ }\href@noop {} {\bibfield
  {journal} {\bibinfo  {journal} {Science advances}\ }\textbf {\bibinfo
  {volume} {6}},\ \bibinfo {pages} {eabb6003} (\bibinfo {year}
  {2020})}\BibitemShut {NoStop}%
\bibitem [{\citenamefont {Nayak}\ \emph {et~al.}(2016)\citenamefont {Nayak},
  \citenamefont {Fischer}, \citenamefont {Sun}, \citenamefont {Yan},
  \citenamefont {Karel}, \citenamefont {Komarek}, \citenamefont {Shekhar},
  \citenamefont {Kumar}, \citenamefont {Schnelle}, \citenamefont {K{\"u}bler}
  \emph {et~al.}}]{nayak2016large}%
  \BibitemOpen
  \bibfield  {author} {\bibinfo {author} {\bibfnamefont {A.~K.}\ \bibnamefont
  {Nayak}}, \bibinfo {author} {\bibfnamefont {J.~E.}\ \bibnamefont {Fischer}},
  \bibinfo {author} {\bibfnamefont {Y.}~\bibnamefont {Sun}}, \bibinfo {author}
  {\bibfnamefont {B.}~\bibnamefont {Yan}}, \bibinfo {author} {\bibfnamefont
  {J.}~\bibnamefont {Karel}}, \bibinfo {author} {\bibfnamefont {A.~C.}\
  \bibnamefont {Komarek}}, \bibinfo {author} {\bibfnamefont {C.}~\bibnamefont
  {Shekhar}}, \bibinfo {author} {\bibfnamefont {N.}~\bibnamefont {Kumar}},
  \bibinfo {author} {\bibfnamefont {W.}~\bibnamefont {Schnelle}}, \bibinfo
  {author} {\bibfnamefont {J.}~\bibnamefont {K{\"u}bler}},  \emph {et~al.},\
  }\href@noop {} {\bibfield  {journal} {\bibinfo  {journal} {Science advances}\
  }\textbf {\bibinfo {volume} {2}},\ \bibinfo {pages} {e1501870} (\bibinfo
  {year} {2016})}\BibitemShut {NoStop}%
\bibitem [{\citenamefont {Kurumaji}\ \emph {et~al.}(2019)\citenamefont
  {Kurumaji}, \citenamefont {Nakajima}, \citenamefont {Hirschberger},
  \citenamefont {Kikkawa}, \citenamefont {Yamasaki}, \citenamefont {Sagayama},
  \citenamefont {Nakao}, \citenamefont {Taguchi}, \citenamefont {Arima},\ and\
  \citenamefont {Tokura}}]{kurumaji2019skyrmion}%
  \BibitemOpen
  \bibfield  {author} {\bibinfo {author} {\bibfnamefont {T.}~\bibnamefont
  {Kurumaji}}, \bibinfo {author} {\bibfnamefont {T.}~\bibnamefont {Nakajima}},
  \bibinfo {author} {\bibfnamefont {M.}~\bibnamefont {Hirschberger}}, \bibinfo
  {author} {\bibfnamefont {A.}~\bibnamefont {Kikkawa}}, \bibinfo {author}
  {\bibfnamefont {Y.}~\bibnamefont {Yamasaki}}, \bibinfo {author}
  {\bibfnamefont {H.}~\bibnamefont {Sagayama}}, \bibinfo {author}
  {\bibfnamefont {H.}~\bibnamefont {Nakao}}, \bibinfo {author} {\bibfnamefont
  {Y.}~\bibnamefont {Taguchi}}, \bibinfo {author} {\bibfnamefont {T.-h.}\
  \bibnamefont {Arima}}, \ and\ \bibinfo {author} {\bibfnamefont
  {Y.}~\bibnamefont {Tokura}},\ }\href@noop {} {\bibfield  {journal} {\bibinfo
  {journal} {Science}\ }\textbf {\bibinfo {volume} {365}},\ \bibinfo {pages}
  {914} (\bibinfo {year} {2019})}\BibitemShut {NoStop}%
\bibitem [{\citenamefont {Fujishiro}\ \emph {et~al.}(2021)\citenamefont
  {Fujishiro}, \citenamefont {Kanazawa}, \citenamefont {Kurihara},
  \citenamefont {Ishizuka}, \citenamefont {Hori}, \citenamefont {Yasin},
  \citenamefont {Yu}, \citenamefont {Tsukazaki}, \citenamefont {Ichikawa},
  \citenamefont {Kawasaki} \emph {et~al.}}]{fujishiro2021giant}%
  \BibitemOpen
  \bibfield  {author} {\bibinfo {author} {\bibfnamefont {Y.}~\bibnamefont
  {Fujishiro}}, \bibinfo {author} {\bibfnamefont {N.}~\bibnamefont {Kanazawa}},
  \bibinfo {author} {\bibfnamefont {R.}~\bibnamefont {Kurihara}}, \bibinfo
  {author} {\bibfnamefont {H.}~\bibnamefont {Ishizuka}}, \bibinfo {author}
  {\bibfnamefont {T.}~\bibnamefont {Hori}}, \bibinfo {author} {\bibfnamefont
  {F.~S.}\ \bibnamefont {Yasin}}, \bibinfo {author} {\bibfnamefont
  {X.}~\bibnamefont {Yu}}, \bibinfo {author} {\bibfnamefont {A.}~\bibnamefont
  {Tsukazaki}}, \bibinfo {author} {\bibfnamefont {M.}~\bibnamefont {Ichikawa}},
  \bibinfo {author} {\bibfnamefont {M.}~\bibnamefont {Kawasaki}},  \emph
  {et~al.},\ }\href@noop {} {\bibfield  {journal} {\bibinfo  {journal} {Nature
  communications}\ }\textbf {\bibinfo {volume} {12}},\ \bibinfo {pages} {1}
  (\bibinfo {year} {2021})}\BibitemShut {NoStop}%
\bibitem [{\citenamefont {Lee}\ \emph {et~al.}(2007)\citenamefont {Lee},
  \citenamefont {Onose}, \citenamefont {Tokura},\ and\ \citenamefont
  {Ong}}]{lee2007hidden}%
  \BibitemOpen
  \bibfield  {author} {\bibinfo {author} {\bibfnamefont {M.}~\bibnamefont
  {Lee}}, \bibinfo {author} {\bibfnamefont {Y.}~\bibnamefont {Onose}}, \bibinfo
  {author} {\bibfnamefont {Y.}~\bibnamefont {Tokura}}, \ and\ \bibinfo {author}
  {\bibfnamefont {N.}~\bibnamefont {Ong}},\ }\href@noop {} {\bibfield
  {journal} {\bibinfo  {journal} {Physical Review B}\ }\textbf {\bibinfo
  {volume} {75}},\ \bibinfo {pages} {172403} (\bibinfo {year}
  {2007})}\BibitemShut {NoStop}%
\bibitem [{\citenamefont {Liang}\ \emph {et~al.}(2018)\citenamefont {Liang},
  \citenamefont {Lin}, \citenamefont {Kushwaha}, \citenamefont {Xing},
  \citenamefont {Ni}, \citenamefont {Cava},\ and\ \citenamefont
  {Ong}}]{liang2018experimental}%
  \BibitemOpen
  \bibfield  {author} {\bibinfo {author} {\bibfnamefont {S.}~\bibnamefont
  {Liang}}, \bibinfo {author} {\bibfnamefont {J.}~\bibnamefont {Lin}}, \bibinfo
  {author} {\bibfnamefont {S.}~\bibnamefont {Kushwaha}}, \bibinfo {author}
  {\bibfnamefont {J.}~\bibnamefont {Xing}}, \bibinfo {author} {\bibfnamefont
  {N.}~\bibnamefont {Ni}}, \bibinfo {author} {\bibfnamefont {R.~J.}\
  \bibnamefont {Cava}}, \ and\ \bibinfo {author} {\bibfnamefont {N.~P.}\
  \bibnamefont {Ong}},\ }\href@noop {} {\bibfield  {journal} {\bibinfo
  {journal} {Physical Review X}\ }\textbf {\bibinfo {volume} {8}},\ \bibinfo
  {pages} {031002} (\bibinfo {year} {2018})}\BibitemShut {NoStop}%
\bibitem [{\citenamefont {Singha}\ \emph {et~al.}(2019)\citenamefont {Singha},
  \citenamefont {Roy}, \citenamefont {Pariari}, \citenamefont {Satpati},\ and\
  \citenamefont {Mandal}}]{singha2019magnetotransport}%
  \BibitemOpen
  \bibfield  {author} {\bibinfo {author} {\bibfnamefont {R.}~\bibnamefont
  {Singha}}, \bibinfo {author} {\bibfnamefont {S.}~\bibnamefont {Roy}},
  \bibinfo {author} {\bibfnamefont {A.}~\bibnamefont {Pariari}}, \bibinfo
  {author} {\bibfnamefont {B.}~\bibnamefont {Satpati}}, \ and\ \bibinfo
  {author} {\bibfnamefont {P.}~\bibnamefont {Mandal}},\ }\href@noop {}
  {\bibfield  {journal} {\bibinfo  {journal} {Physical Review B}\ }\textbf
  {\bibinfo {volume} {99}},\ \bibinfo {pages} {035110} (\bibinfo {year}
  {2019})}\BibitemShut {NoStop}%
\bibitem [{\citenamefont {Chen}\ \emph {et~al.}(2021)\citenamefont {Chen},
  \citenamefont {Li}, \citenamefont {Ding}, \citenamefont {Zhang},
  \citenamefont {Liu},\ and\ \citenamefont {Wang}}]{chen2021large}%
  \BibitemOpen
  \bibfield  {author} {\bibinfo {author} {\bibfnamefont {J.}~\bibnamefont
  {Chen}}, \bibinfo {author} {\bibfnamefont {H.}~\bibnamefont {Li}}, \bibinfo
  {author} {\bibfnamefont {B.}~\bibnamefont {Ding}}, \bibinfo {author}
  {\bibfnamefont {H.}~\bibnamefont {Zhang}}, \bibinfo {author} {\bibfnamefont
  {E.}~\bibnamefont {Liu}}, \ and\ \bibinfo {author} {\bibfnamefont
  {W.}~\bibnamefont {Wang}},\ }\href@noop {} {\bibfield  {journal} {\bibinfo
  {journal} {Applied Physics Letters}\ }\textbf {\bibinfo {volume} {118}},\
  \bibinfo {pages} {031901} (\bibinfo {year} {2021})}\BibitemShut {NoStop}%
\bibitem [{\citenamefont {Pavlosiuk}\ \emph {et~al.}(2020)\citenamefont
  {Pavlosiuk}, \citenamefont {Fa{\l}at}, \citenamefont {Kaczorowski},\ and\
  \citenamefont {Wi{\'s}niewski}}]{pavlosiuk2020anomalous}%
  \BibitemOpen
  \bibfield  {author} {\bibinfo {author} {\bibfnamefont {O.}~\bibnamefont
  {Pavlosiuk}}, \bibinfo {author} {\bibfnamefont {P.}~\bibnamefont {Fa{\l}at}},
  \bibinfo {author} {\bibfnamefont {D.}~\bibnamefont {Kaczorowski}}, \ and\
  \bibinfo {author} {\bibfnamefont {P.}~\bibnamefont {Wi{\'s}niewski}},\
  }\href@noop {} {\bibfield  {journal} {\bibinfo  {journal} {APL Materials}\
  }\textbf {\bibinfo {volume} {8}},\ \bibinfo {pages} {111107} (\bibinfo {year}
  {2020})}\BibitemShut {NoStop}%
\bibitem [{\citenamefont {Zhu}\ \emph {et~al.}(2020)\citenamefont {Zhu},
  \citenamefont {Singh}, \citenamefont {Wang}, \citenamefont {Huang},
  \citenamefont {Chiu}, \citenamefont {Wang}, \citenamefont {Graf},
  \citenamefont {Zhang}, \citenamefont {Lin}, \citenamefont {Sun} \emph
  {et~al.}}]{zhu2020exceptionally}%
  \BibitemOpen
  \bibfield  {author} {\bibinfo {author} {\bibfnamefont {Y.}~\bibnamefont
  {Zhu}}, \bibinfo {author} {\bibfnamefont {B.}~\bibnamefont {Singh}}, \bibinfo
  {author} {\bibfnamefont {Y.}~\bibnamefont {Wang}}, \bibinfo {author}
  {\bibfnamefont {C.-Y.}\ \bibnamefont {Huang}}, \bibinfo {author}
  {\bibfnamefont {W.-C.}\ \bibnamefont {Chiu}}, \bibinfo {author}
  {\bibfnamefont {B.}~\bibnamefont {Wang}}, \bibinfo {author} {\bibfnamefont
  {D.}~\bibnamefont {Graf}}, \bibinfo {author} {\bibfnamefont {Y.}~\bibnamefont
  {Zhang}}, \bibinfo {author} {\bibfnamefont {H.}~\bibnamefont {Lin}}, \bibinfo
  {author} {\bibfnamefont {J.}~\bibnamefont {Sun}},  \emph {et~al.},\
  }\href@noop {} {\bibfield  {journal} {\bibinfo  {journal} {Physical Review
  B}\ }\textbf {\bibinfo {volume} {101}},\ \bibinfo {pages} {161105} (\bibinfo
  {year} {2020})}\BibitemShut {NoStop}%
\bibitem [{\citenamefont {Lee}\ \emph {et~al.}(2009)\citenamefont {Lee},
  \citenamefont {Kang}, \citenamefont {Onose}, \citenamefont {Tokura},\ and\
  \citenamefont {Ong}}]{lee2009unusual}%
  \BibitemOpen
  \bibfield  {author} {\bibinfo {author} {\bibfnamefont {M.}~\bibnamefont
  {Lee}}, \bibinfo {author} {\bibfnamefont {W.}~\bibnamefont {Kang}}, \bibinfo
  {author} {\bibfnamefont {Y.}~\bibnamefont {Onose}}, \bibinfo {author}
  {\bibfnamefont {Y.}~\bibnamefont {Tokura}}, \ and\ \bibinfo {author}
  {\bibfnamefont {N.~P.}\ \bibnamefont {Ong}},\ }\href@noop {} {\bibfield
  {journal} {\bibinfo  {journal} {Physical review letters}\ }\textbf {\bibinfo
  {volume} {102}},\ \bibinfo {pages} {186601} (\bibinfo {year}
  {2009})}\BibitemShut {NoStop}%
\bibitem [{\citenamefont {Nagaosa}\ \emph {et~al.}(2010)\citenamefont
  {Nagaosa}, \citenamefont {Sinova}, \citenamefont {Onoda}, \citenamefont
  {MacDonald},\ and\ \citenamefont {Ong}}]{nagaosa2010anomalous}%
  \BibitemOpen
  \bibfield  {author} {\bibinfo {author} {\bibfnamefont {N.}~\bibnamefont
  {Nagaosa}}, \bibinfo {author} {\bibfnamefont {J.}~\bibnamefont {Sinova}},
  \bibinfo {author} {\bibfnamefont {S.}~\bibnamefont {Onoda}}, \bibinfo
  {author} {\bibfnamefont {A.~H.}\ \bibnamefont {MacDonald}}, \ and\ \bibinfo
  {author} {\bibfnamefont {N.~P.}\ \bibnamefont {Ong}},\ }\href@noop {}
  {\bibfield  {journal} {\bibinfo  {journal} {Reviews of modern physics}\
  }\textbf {\bibinfo {volume} {82}},\ \bibinfo {pages} {1539} (\bibinfo {year}
  {2010})}\BibitemShut {NoStop}%
\bibitem [{\citenamefont {Xiao}\ \emph {et~al.}(2010)\citenamefont {Xiao},
  \citenamefont {Chang},\ and\ \citenamefont {Niu}}]{xiao2010berry}%
  \BibitemOpen
  \bibfield  {author} {\bibinfo {author} {\bibfnamefont {D.}~\bibnamefont
  {Xiao}}, \bibinfo {author} {\bibfnamefont {M.-C.}\ \bibnamefont {Chang}}, \
  and\ \bibinfo {author} {\bibfnamefont {Q.}~\bibnamefont {Niu}},\ }\href@noop
  {} {\bibfield  {journal} {\bibinfo  {journal} {Reviews of modern physics}\
  }\textbf {\bibinfo {volume} {82}},\ \bibinfo {pages} {1959} (\bibinfo {year}
  {2010})}\BibitemShut {NoStop}%
\bibitem [{\citenamefont {Karplus}\ and\ \citenamefont
  {Luttinger}(1954)}]{karplus1954hall}%
  \BibitemOpen
  \bibfield  {author} {\bibinfo {author} {\bibfnamefont {R.}~\bibnamefont
  {Karplus}}\ and\ \bibinfo {author} {\bibfnamefont {J.}~\bibnamefont
  {Luttinger}},\ }\href@noop {} {\bibfield  {journal} {\bibinfo  {journal}
  {Physical Review}\ }\textbf {\bibinfo {volume} {95}},\ \bibinfo {pages}
  {1154} (\bibinfo {year} {1954})}\BibitemShut {NoStop}%
\bibitem [{\citenamefont {Ohgushi}\ \emph {et~al.}(2000)\citenamefont
  {Ohgushi}, \citenamefont {Murakami},\ and\ \citenamefont
  {Nagaosa}}]{ohgushi2000spin}%
  \BibitemOpen
  \bibfield  {author} {\bibinfo {author} {\bibfnamefont {K.}~\bibnamefont
  {Ohgushi}}, \bibinfo {author} {\bibfnamefont {S.}~\bibnamefont {Murakami}}, \
  and\ \bibinfo {author} {\bibfnamefont {N.}~\bibnamefont {Nagaosa}},\
  }\href@noop {} {\bibfield  {journal} {\bibinfo  {journal} {Physical Review
  B}\ }\textbf {\bibinfo {volume} {62}},\ \bibinfo {pages} {R6065} (\bibinfo
  {year} {2000})}\BibitemShut {NoStop}%
\bibitem [{\citenamefont {Shindou}\ and\ \citenamefont
  {Nagaosa}(2001)}]{shindou2001orbital}%
  \BibitemOpen
  \bibfield  {author} {\bibinfo {author} {\bibfnamefont {R.}~\bibnamefont
  {Shindou}}\ and\ \bibinfo {author} {\bibfnamefont {N.}~\bibnamefont
  {Nagaosa}},\ }\href@noop {} {\bibfield  {journal} {\bibinfo  {journal}
  {Physical review letters}\ }\textbf {\bibinfo {volume} {87}},\ \bibinfo
  {pages} {116801} (\bibinfo {year} {2001})}\BibitemShut {NoStop}%
\bibitem [{\citenamefont {Martin}\ and\ \citenamefont
  {Batista}(2008)}]{martin2008itinerant}%
  \BibitemOpen
  \bibfield  {author} {\bibinfo {author} {\bibfnamefont {I.}~\bibnamefont
  {Martin}}\ and\ \bibinfo {author} {\bibfnamefont {C.}~\bibnamefont
  {Batista}},\ }\href@noop {} {\bibfield  {journal} {\bibinfo  {journal}
  {Physical review letters}\ }\textbf {\bibinfo {volume} {101}},\ \bibinfo
  {pages} {156402} (\bibinfo {year} {2008})}\BibitemShut {NoStop}%
\bibitem [{\citenamefont {Thouless}\ \emph {et~al.}(1982)\citenamefont
  {Thouless}, \citenamefont {Kohmoto}, \citenamefont {Nightingale},\ and\
  \citenamefont {den Nijs}}]{thouless1982quantized}%
  \BibitemOpen
  \bibfield  {author} {\bibinfo {author} {\bibfnamefont {D.~J.}\ \bibnamefont
  {Thouless}}, \bibinfo {author} {\bibfnamefont {M.}~\bibnamefont {Kohmoto}},
  \bibinfo {author} {\bibfnamefont {M.~P.}\ \bibnamefont {Nightingale}}, \ and\
  \bibinfo {author} {\bibfnamefont {M.}~\bibnamefont {den Nijs}},\ }\href@noop
  {} {\bibfield  {journal} {\bibinfo  {journal} {Physical review letters}\
  }\textbf {\bibinfo {volume} {49}},\ \bibinfo {pages} {405} (\bibinfo {year}
  {1982})}\BibitemShut {NoStop}%
\bibitem [{\citenamefont {Haldane}(2004)}]{haldane2004berry}%
  \BibitemOpen
  \bibfield  {author} {\bibinfo {author} {\bibfnamefont {F.}~\bibnamefont
  {Haldane}},\ }\href@noop {} {\bibfield  {journal} {\bibinfo  {journal}
  {Physical review letters}\ }\textbf {\bibinfo {volume} {93}},\ \bibinfo
  {pages} {206602} (\bibinfo {year} {2004})}\BibitemShut {NoStop}%
\bibitem [{\citenamefont {Hirschberger}\ \emph {et~al.}(2019)\citenamefont
  {Hirschberger}, \citenamefont {Nakajima}, \citenamefont {Gao}, \citenamefont
  {Peng}, \citenamefont {Kikkawa}, \citenamefont {Kurumaji}, \citenamefont
  {Kriener}, \citenamefont {Yamasaki}, \citenamefont {Sagayama}, \citenamefont
  {Nakao} \emph {et~al.}}]{hirschberger2019skyrmion}%
  \BibitemOpen
  \bibfield  {author} {\bibinfo {author} {\bibfnamefont {M.}~\bibnamefont
  {Hirschberger}}, \bibinfo {author} {\bibfnamefont {T.}~\bibnamefont
  {Nakajima}}, \bibinfo {author} {\bibfnamefont {S.}~\bibnamefont {Gao}},
  \bibinfo {author} {\bibfnamefont {L.}~\bibnamefont {Peng}}, \bibinfo {author}
  {\bibfnamefont {A.}~\bibnamefont {Kikkawa}}, \bibinfo {author} {\bibfnamefont
  {T.}~\bibnamefont {Kurumaji}}, \bibinfo {author} {\bibfnamefont
  {M.}~\bibnamefont {Kriener}}, \bibinfo {author} {\bibfnamefont
  {Y.}~\bibnamefont {Yamasaki}}, \bibinfo {author} {\bibfnamefont
  {H.}~\bibnamefont {Sagayama}}, \bibinfo {author} {\bibfnamefont
  {H.}~\bibnamefont {Nakao}},  \emph {et~al.},\ }\href@noop {} {\bibfield
  {journal} {\bibinfo  {journal} {Nature communications}\ }\textbf {\bibinfo
  {volume} {10}},\ \bibinfo {pages} {1} (\bibinfo {year} {2019})}\BibitemShut
  {NoStop}%
\bibitem [{\citenamefont {Nakamura}\ \emph {et~al.}(2013)\citenamefont
  {Nakamura}, \citenamefont {Hiranaka}, \citenamefont {Hedo}, \citenamefont
  {Nakama}, \citenamefont {Tatetsu}, \citenamefont {Maehira}, \citenamefont
  {Miura}, \citenamefont {Mori}, \citenamefont {Tsutsumi}, \citenamefont
  {Hirose} \emph {et~al.}}]{nakamura2013fermi}%
  \BibitemOpen
  \bibfield  {author} {\bibinfo {author} {\bibfnamefont {A.}~\bibnamefont
  {Nakamura}}, \bibinfo {author} {\bibfnamefont {Y.}~\bibnamefont {Hiranaka}},
  \bibinfo {author} {\bibfnamefont {M.}~\bibnamefont {Hedo}}, \bibinfo {author}
  {\bibfnamefont {T.}~\bibnamefont {Nakama}}, \bibinfo {author} {\bibfnamefont
  {Y.}~\bibnamefont {Tatetsu}}, \bibinfo {author} {\bibfnamefont
  {T.}~\bibnamefont {Maehira}}, \bibinfo {author} {\bibfnamefont
  {Y.}~\bibnamefont {Miura}}, \bibinfo {author} {\bibfnamefont
  {A.}~\bibnamefont {Mori}}, \bibinfo {author} {\bibfnamefont {H.}~\bibnamefont
  {Tsutsumi}}, \bibinfo {author} {\bibfnamefont {Y.}~\bibnamefont {Hirose}},
  \emph {et~al.},\ }\href@noop {} {\bibfield  {journal} {\bibinfo  {journal}
  {Journal of the Physical Society of Japan}\ }\textbf {\bibinfo {volume}
  {82}},\ \bibinfo {pages} {124708} (\bibinfo {year} {2013})}\BibitemShut
  {NoStop}%
\bibitem [{\citenamefont {Neubauer}\ \emph {et~al.}(2009)\citenamefont
  {Neubauer}, \citenamefont {Pfleiderer}, \citenamefont {Binz}, \citenamefont
  {Rosch}, \citenamefont {Ritz}, \citenamefont {Niklowitz},\ and\ \citenamefont
  {B{\"o}ni}}]{neubauer2009topological}%
  \BibitemOpen
  \bibfield  {author} {\bibinfo {author} {\bibfnamefont {A.}~\bibnamefont
  {Neubauer}}, \bibinfo {author} {\bibfnamefont {C.}~\bibnamefont
  {Pfleiderer}}, \bibinfo {author} {\bibfnamefont {B.}~\bibnamefont {Binz}},
  \bibinfo {author} {\bibfnamefont {A.}~\bibnamefont {Rosch}}, \bibinfo
  {author} {\bibfnamefont {R.}~\bibnamefont {Ritz}}, \bibinfo {author}
  {\bibfnamefont {P.}~\bibnamefont {Niklowitz}}, \ and\ \bibinfo {author}
  {\bibfnamefont {P.}~\bibnamefont {B{\"o}ni}},\ }\href@noop {} {\bibfield
  {journal} {\bibinfo  {journal} {Physical review letters}\ }\textbf {\bibinfo
  {volume} {102}},\ \bibinfo {pages} {186602} (\bibinfo {year}
  {2009})}\BibitemShut {NoStop}%
\bibitem [{\citenamefont {Khanh}\ \emph {et~al.}(2020)\citenamefont {Khanh},
  \citenamefont {Nakajima}, \citenamefont {Yu}, \citenamefont {Gao},
  \citenamefont {Shibata}, \citenamefont {Hirschberger}, \citenamefont
  {Yamasaki}, \citenamefont {Sagayama}, \citenamefont {Nakao}, \citenamefont
  {Peng} \emph {et~al.}}]{khanh2020nanometric}%
  \BibitemOpen
  \bibfield  {author} {\bibinfo {author} {\bibfnamefont {N.~D.}\ \bibnamefont
  {Khanh}}, \bibinfo {author} {\bibfnamefont {T.}~\bibnamefont {Nakajima}},
  \bibinfo {author} {\bibfnamefont {X.}~\bibnamefont {Yu}}, \bibinfo {author}
  {\bibfnamefont {S.}~\bibnamefont {Gao}}, \bibinfo {author} {\bibfnamefont
  {K.}~\bibnamefont {Shibata}}, \bibinfo {author} {\bibfnamefont
  {M.}~\bibnamefont {Hirschberger}}, \bibinfo {author} {\bibfnamefont
  {Y.}~\bibnamefont {Yamasaki}}, \bibinfo {author} {\bibfnamefont
  {H.}~\bibnamefont {Sagayama}}, \bibinfo {author} {\bibfnamefont
  {H.}~\bibnamefont {Nakao}}, \bibinfo {author} {\bibfnamefont
  {L.}~\bibnamefont {Peng}},  \emph {et~al.},\ }\href@noop {} {\bibfield
  {journal} {\bibinfo  {journal} {Nature Nanotechnology}\ }\textbf {\bibinfo
  {volume} {15}},\ \bibinfo {pages} {444} (\bibinfo {year} {2020})}\BibitemShut
  {NoStop}%
\bibitem [{\citenamefont {Onoda}\ \emph {et~al.}(2006)\citenamefont {Onoda},
  \citenamefont {Sugimoto},\ and\ \citenamefont
  {Nagaosa}}]{onoda2006intrinsic}%
  \BibitemOpen
  \bibfield  {author} {\bibinfo {author} {\bibfnamefont {S.}~\bibnamefont
  {Onoda}}, \bibinfo {author} {\bibfnamefont {N.}~\bibnamefont {Sugimoto}}, \
  and\ \bibinfo {author} {\bibfnamefont {N.}~\bibnamefont {Nagaosa}},\
  }\href@noop {} {\bibfield  {journal} {\bibinfo  {journal} {Physical review
  letters}\ }\textbf {\bibinfo {volume} {97}},\ \bibinfo {pages} {126602}
  (\bibinfo {year} {2006})}\BibitemShut {NoStop}%
\bibitem [{\citenamefont {Wang}\ \emph {et~al.}(2021)\citenamefont {Wang},
  \citenamefont {Neubauer}, \citenamefont {Duan}, \citenamefont {Yin},
  \citenamefont {Fujitsu}, \citenamefont {Hosono}, \citenamefont {Ye},
  \citenamefont {Zhang}, \citenamefont {Chi}, \citenamefont {Krycka} \emph
  {et~al.}}]{wang2021field}%
  \BibitemOpen
  \bibfield  {author} {\bibinfo {author} {\bibfnamefont {Q.}~\bibnamefont
  {Wang}}, \bibinfo {author} {\bibfnamefont {K.~J.}\ \bibnamefont {Neubauer}},
  \bibinfo {author} {\bibfnamefont {C.}~\bibnamefont {Duan}}, \bibinfo {author}
  {\bibfnamefont {Q.}~\bibnamefont {Yin}}, \bibinfo {author} {\bibfnamefont
  {S.}~\bibnamefont {Fujitsu}}, \bibinfo {author} {\bibfnamefont
  {H.}~\bibnamefont {Hosono}}, \bibinfo {author} {\bibfnamefont
  {F.}~\bibnamefont {Ye}}, \bibinfo {author} {\bibfnamefont {R.}~\bibnamefont
  {Zhang}}, \bibinfo {author} {\bibfnamefont {S.}~\bibnamefont {Chi}}, \bibinfo
  {author} {\bibfnamefont {K.}~\bibnamefont {Krycka}},  \emph {et~al.},\
  }\href@noop {} {\bibfield  {journal} {\bibinfo  {journal} {Physical Review
  B}\ }\textbf {\bibinfo {volume} {103}},\ \bibinfo {pages} {014416} (\bibinfo
  {year} {2021})}\BibitemShut {NoStop}%
\bibitem [{\citenamefont {Ghimire}\ \emph {et~al.}(2020)\citenamefont
  {Ghimire}, \citenamefont {Dally}, \citenamefont {Poudel}, \citenamefont
  {Jones}, \citenamefont {Michel}, \citenamefont {Magar}, \citenamefont
  {Bleuel}, \citenamefont {McGuire}, \citenamefont {Jiang}, \citenamefont
  {Mitchell} \emph {et~al.}}]{ghimire2020competing}%
  \BibitemOpen
  \bibfield  {author} {\bibinfo {author} {\bibfnamefont {N.~J.}\ \bibnamefont
  {Ghimire}}, \bibinfo {author} {\bibfnamefont {R.~L.}\ \bibnamefont {Dally}},
  \bibinfo {author} {\bibfnamefont {L.}~\bibnamefont {Poudel}}, \bibinfo
  {author} {\bibfnamefont {D.}~\bibnamefont {Jones}}, \bibinfo {author}
  {\bibfnamefont {D.}~\bibnamefont {Michel}}, \bibinfo {author} {\bibfnamefont
  {N.~T.}\ \bibnamefont {Magar}}, \bibinfo {author} {\bibfnamefont
  {M.}~\bibnamefont {Bleuel}}, \bibinfo {author} {\bibfnamefont {M.~A.}\
  \bibnamefont {McGuire}}, \bibinfo {author} {\bibfnamefont {J.}~\bibnamefont
  {Jiang}}, \bibinfo {author} {\bibfnamefont {J.}~\bibnamefont {Mitchell}},
  \emph {et~al.},\ }\href@noop {} {\bibfield  {journal} {\bibinfo  {journal}
  {Science Advances}\ }\textbf {\bibinfo {volume} {6}},\ \bibinfo {pages}
  {eabe2680} (\bibinfo {year} {2020})}\BibitemShut {NoStop}%
\bibitem [{\citenamefont
  {Dzyaloshinsky}(1958)}]{dzyaloshinsky1958thermodynamic}%
  \BibitemOpen
  \bibfield  {author} {\bibinfo {author} {\bibfnamefont {I.}~\bibnamefont
  {Dzyaloshinsky}},\ }\href@noop {} {\bibfield  {journal} {\bibinfo  {journal}
  {Journal of Physics and Chemistry of Solids}\ }\textbf {\bibinfo {volume}
  {4}},\ \bibinfo {pages} {241} (\bibinfo {year} {1958})}\BibitemShut {NoStop}%
\bibitem [{\citenamefont {Moriya}(1960)}]{moriya1960new}%
  \BibitemOpen
  \bibfield  {author} {\bibinfo {author} {\bibfnamefont {T.}~\bibnamefont
  {Moriya}},\ }\href@noop {} {\bibfield  {journal} {\bibinfo  {journal}
  {Physical Review Letters}\ }\textbf {\bibinfo {volume} {4}},\ \bibinfo
  {pages} {228} (\bibinfo {year} {1960})}\BibitemShut {NoStop}%
\bibitem [{\citenamefont {Ozawa}\ \emph {et~al.}(2017)\citenamefont {Ozawa},
  \citenamefont {Hayami},\ and\ \citenamefont {Motome}}]{ozawa2017zero}%
  \BibitemOpen
  \bibfield  {author} {\bibinfo {author} {\bibfnamefont {R.}~\bibnamefont
  {Ozawa}}, \bibinfo {author} {\bibfnamefont {S.}~\bibnamefont {Hayami}}, \
  and\ \bibinfo {author} {\bibfnamefont {Y.}~\bibnamefont {Motome}},\
  }\href@noop {} {\bibfield  {journal} {\bibinfo  {journal} {Physical Review
  Letters}\ }\textbf {\bibinfo {volume} {118}},\ \bibinfo {pages} {147205}
  (\bibinfo {year} {2017})}\BibitemShut {NoStop}%
\bibitem [{\citenamefont {Hayami}\ \emph {et~al.}(2017)\citenamefont {Hayami},
  \citenamefont {Ozawa},\ and\ \citenamefont {Motome}}]{hayami2017effective}%
  \BibitemOpen
  \bibfield  {author} {\bibinfo {author} {\bibfnamefont {S.}~\bibnamefont
  {Hayami}}, \bibinfo {author} {\bibfnamefont {R.}~\bibnamefont {Ozawa}}, \
  and\ \bibinfo {author} {\bibfnamefont {Y.}~\bibnamefont {Motome}},\
  }\href@noop {} {\bibfield  {journal} {\bibinfo  {journal} {Physical Review
  B}\ }\textbf {\bibinfo {volume} {95}},\ \bibinfo {pages} {224424} (\bibinfo
  {year} {2017})}\BibitemShut {NoStop}%
\bibitem [{\citenamefont {Wang}\ \emph
  {et~al.}(2020{\natexlab{a}})\citenamefont {Wang}, \citenamefont {Su},
  \citenamefont {Lin},\ and\ \citenamefont {Batista}}]{wang2020skyrmion}%
  \BibitemOpen
  \bibfield  {author} {\bibinfo {author} {\bibfnamefont {Z.}~\bibnamefont
  {Wang}}, \bibinfo {author} {\bibfnamefont {Y.}~\bibnamefont {Su}}, \bibinfo
  {author} {\bibfnamefont {S.-Z.}\ \bibnamefont {Lin}}, \ and\ \bibinfo
  {author} {\bibfnamefont {C.~D.}\ \bibnamefont {Batista}},\ }\href@noop {}
  {\bibfield  {journal} {\bibinfo  {journal} {Physical Review Letters}\
  }\textbf {\bibinfo {volume} {124}},\ \bibinfo {pages} {207201} (\bibinfo
  {year} {2020}{\natexlab{a}})}\BibitemShut {NoStop}%
\bibitem [{\citenamefont {Hayami}\ and\ \citenamefont
  {Motome}(2021)}]{hayami2021square}%
  \BibitemOpen
  \bibfield  {author} {\bibinfo {author} {\bibfnamefont {S.}~\bibnamefont
  {Hayami}}\ and\ \bibinfo {author} {\bibfnamefont {Y.}~\bibnamefont
  {Motome}},\ }\href@noop {} {\bibfield  {journal} {\bibinfo  {journal}
  {Physical Review B}\ }\textbf {\bibinfo {volume} {103}},\ \bibinfo {pages}
  {024439} (\bibinfo {year} {2021})}\BibitemShut {NoStop}%
\bibitem [{\citenamefont {Ruderman}\ and\ \citenamefont
  {Kittel}(1954)}]{ruderman1954indirect}%
  \BibitemOpen
  \bibfield  {author} {\bibinfo {author} {\bibfnamefont {M.~A.}\ \bibnamefont
  {Ruderman}}\ and\ \bibinfo {author} {\bibfnamefont {C.}~\bibnamefont
  {Kittel}},\ }\href@noop {} {\bibfield  {journal} {\bibinfo  {journal}
  {Physical Review}\ }\textbf {\bibinfo {volume} {96}},\ \bibinfo {pages} {99}
  (\bibinfo {year} {1954})}\BibitemShut {NoStop}%
\bibitem [{\citenamefont {Kasuya}(1956)}]{kasuya1956theory}%
  \BibitemOpen
  \bibfield  {author} {\bibinfo {author} {\bibfnamefont {T.}~\bibnamefont
  {Kasuya}},\ }\href@noop {} {\bibfield  {journal} {\bibinfo  {journal}
  {Progress of theoretical physics}\ }\textbf {\bibinfo {volume} {16}},\
  \bibinfo {pages} {45} (\bibinfo {year} {1956})}\BibitemShut {NoStop}%
\bibitem [{\citenamefont {Yosida}(1957)}]{yosida1957magnetic}%
  \BibitemOpen
  \bibfield  {author} {\bibinfo {author} {\bibfnamefont {K.}~\bibnamefont
  {Yosida}},\ }\href@noop {} {\bibfield  {journal} {\bibinfo  {journal}
  {Physical Review}\ }\textbf {\bibinfo {volume} {106}},\ \bibinfo {pages}
  {893} (\bibinfo {year} {1957})}\BibitemShut {NoStop}%
\bibitem [{\citenamefont {Takahashi}\ \emph {et~al.}(2018)\citenamefont
  {Takahashi}, \citenamefont {Ishizuka}, \citenamefont {Murata}, \citenamefont
  {Wang}, \citenamefont {Tokura}, \citenamefont {Nagaosa},\ and\ \citenamefont
  {Kawasaki}}]{takahashi2018anomalous}%
  \BibitemOpen
  \bibfield  {author} {\bibinfo {author} {\bibfnamefont {K.~S.}\ \bibnamefont
  {Takahashi}}, \bibinfo {author} {\bibfnamefont {H.}~\bibnamefont {Ishizuka}},
  \bibinfo {author} {\bibfnamefont {T.}~\bibnamefont {Murata}}, \bibinfo
  {author} {\bibfnamefont {Q.~Y.}\ \bibnamefont {Wang}}, \bibinfo {author}
  {\bibfnamefont {Y.}~\bibnamefont {Tokura}}, \bibinfo {author} {\bibfnamefont
  {N.}~\bibnamefont {Nagaosa}}, \ and\ \bibinfo {author} {\bibfnamefont
  {M.}~\bibnamefont {Kawasaki}},\ }\href@noop {} {\bibfield  {journal}
  {\bibinfo  {journal} {Science advances}\ }\textbf {\bibinfo {volume} {4}},\
  \bibinfo {pages} {eaar7880} (\bibinfo {year} {2018})}\BibitemShut {NoStop}%
\bibitem [{\citenamefont {Miyasato}\ \emph {et~al.}(2007)\citenamefont
  {Miyasato}, \citenamefont {Abe}, \citenamefont {Fujii}, \citenamefont
  {Asamitsu}, \citenamefont {Onoda}, \citenamefont {Onose}, \citenamefont
  {Nagaosa},\ and\ \citenamefont {Tokura}}]{miyasato2007crossover}%
  \BibitemOpen
  \bibfield  {author} {\bibinfo {author} {\bibfnamefont {T.}~\bibnamefont
  {Miyasato}}, \bibinfo {author} {\bibfnamefont {N.}~\bibnamefont {Abe}},
  \bibinfo {author} {\bibfnamefont {T.}~\bibnamefont {Fujii}}, \bibinfo
  {author} {\bibfnamefont {A.}~\bibnamefont {Asamitsu}}, \bibinfo {author}
  {\bibfnamefont {S.}~\bibnamefont {Onoda}}, \bibinfo {author} {\bibfnamefont
  {Y.}~\bibnamefont {Onose}}, \bibinfo {author} {\bibfnamefont
  {N.}~\bibnamefont {Nagaosa}}, \ and\ \bibinfo {author} {\bibfnamefont
  {Y.}~\bibnamefont {Tokura}},\ }\href@noop {} {\bibfield  {journal} {\bibinfo
  {journal} {Physical review letters}\ }\textbf {\bibinfo {volume} {99}},\
  \bibinfo {pages} {086602} (\bibinfo {year} {2007})}\BibitemShut {NoStop}%
\bibitem [{\citenamefont {Iguchi}\ \emph {et~al.}(2007)\citenamefont {Iguchi},
  \citenamefont {Hanasaki},\ and\ \citenamefont {Tokura}}]{iguchi2007scaling}%
  \BibitemOpen
  \bibfield  {author} {\bibinfo {author} {\bibfnamefont {S.}~\bibnamefont
  {Iguchi}}, \bibinfo {author} {\bibfnamefont {N.}~\bibnamefont {Hanasaki}}, \
  and\ \bibinfo {author} {\bibfnamefont {Y.}~\bibnamefont {Tokura}},\
  }\href@noop {} {\bibfield  {journal} {\bibinfo  {journal} {Physical review
  letters}\ }\textbf {\bibinfo {volume} {99}},\ \bibinfo {pages} {077202}
  (\bibinfo {year} {2007})}\BibitemShut {NoStop}%
\bibitem [{\citenamefont {Moya}\ and\ \citenamefont {et~al}(2023)}]{Jaime2023}%
  \BibitemOpen
  \bibfield  {author} {\bibinfo {author} {\bibfnamefont {J.}~\bibnamefont
  {Moya}}\ and\ \bibinfo {author} {\bibnamefont {et~al}},\ }\href@noop {}
  {\bibfield  {journal} {\bibinfo  {journal} {In Preparation}\ } (\bibinfo
  {year} {2023})}\BibitemShut {NoStop}%
\bibitem [{\citenamefont {Wang}\ \emph
  {et~al.}(2020{\natexlab{b}})\citenamefont {Wang}, \citenamefont {Mori},
  \citenamefont {Wang}, \citenamefont {Wang}, \citenamefont {Ma}, \citenamefont
  {Latzke}, \citenamefont {Graf}, \citenamefont {Denlinger}, \citenamefont
  {Campbell}, \citenamefont {Bernevig} \emph {et~al.}}]{wang2020crystalline}%
  \BibitemOpen
  \bibfield  {author} {\bibinfo {author} {\bibfnamefont {K.}~\bibnamefont
  {Wang}}, \bibinfo {author} {\bibfnamefont {R.}~\bibnamefont {Mori}}, \bibinfo
  {author} {\bibfnamefont {Z.}~\bibnamefont {Wang}}, \bibinfo {author}
  {\bibfnamefont {L.}~\bibnamefont {Wang}}, \bibinfo {author} {\bibfnamefont
  {J.~H.~S.}\ \bibnamefont {Ma}}, \bibinfo {author} {\bibfnamefont {D.~W.}\
  \bibnamefont {Latzke}}, \bibinfo {author} {\bibfnamefont {D.~E.}\
  \bibnamefont {Graf}}, \bibinfo {author} {\bibfnamefont {J.~D.}\ \bibnamefont
  {Denlinger}}, \bibinfo {author} {\bibfnamefont {D.}~\bibnamefont {Campbell}},
  \bibinfo {author} {\bibfnamefont {B.~A.}\ \bibnamefont {Bernevig}},  \emph
  {et~al.},\ }\href@noop {} {\bibfield  {journal} {\bibinfo  {journal} {arXiv
  preprint arXiv:2007.12571}\ } (\bibinfo {year}
  {2020}{\natexlab{b}})}\BibitemShut {NoStop}%
\bibitem [{\citenamefont {Hirschberger}\ \emph {et~al.}(2016)\citenamefont
  {Hirschberger}, \citenamefont {Kushwaha}, \citenamefont {Wang}, \citenamefont
  {Gibson}, \citenamefont {Liang}, \citenamefont {Belvin}, \citenamefont
  {Bernevig}, \citenamefont {Cava},\ and\ \citenamefont
  {Ong}}]{hirschberger2016chiral}%
  \BibitemOpen
  \bibfield  {author} {\bibinfo {author} {\bibfnamefont {M.}~\bibnamefont
  {Hirschberger}}, \bibinfo {author} {\bibfnamefont {S.}~\bibnamefont
  {Kushwaha}}, \bibinfo {author} {\bibfnamefont {Z.}~\bibnamefont {Wang}},
  \bibinfo {author} {\bibfnamefont {Q.}~\bibnamefont {Gibson}}, \bibinfo
  {author} {\bibfnamefont {S.}~\bibnamefont {Liang}}, \bibinfo {author}
  {\bibfnamefont {C.~A.}\ \bibnamefont {Belvin}}, \bibinfo {author}
  {\bibfnamefont {B.~A.}\ \bibnamefont {Bernevig}}, \bibinfo {author}
  {\bibfnamefont {R.~J.}\ \bibnamefont {Cava}}, \ and\ \bibinfo {author}
  {\bibfnamefont {N.~P.}\ \bibnamefont {Ong}},\ }\href@noop {} {\bibfield
  {journal} {\bibinfo  {journal} {Nature materials}\ }\textbf {\bibinfo
  {volume} {15}},\ \bibinfo {pages} {1161} (\bibinfo {year}
  {2016})}\BibitemShut {NoStop}%
\bibitem [{\citenamefont {Shekhar}\ \emph {et~al.}(2018)\citenamefont
  {Shekhar}, \citenamefont {Kumar}, \citenamefont {Grinenko}, \citenamefont
  {Singh}, \citenamefont {Sarkar}, \citenamefont {Luetkens}, \citenamefont
  {Wu}, \citenamefont {Zhang}, \citenamefont {Komarek}, \citenamefont {Kampert}
  \emph {et~al.}}]{shekhar2018anomalous}%
  \BibitemOpen
  \bibfield  {author} {\bibinfo {author} {\bibfnamefont {C.}~\bibnamefont
  {Shekhar}}, \bibinfo {author} {\bibfnamefont {N.}~\bibnamefont {Kumar}},
  \bibinfo {author} {\bibfnamefont {V.}~\bibnamefont {Grinenko}}, \bibinfo
  {author} {\bibfnamefont {S.}~\bibnamefont {Singh}}, \bibinfo {author}
  {\bibfnamefont {R.}~\bibnamefont {Sarkar}}, \bibinfo {author} {\bibfnamefont
  {H.}~\bibnamefont {Luetkens}}, \bibinfo {author} {\bibfnamefont {S.-C.}\
  \bibnamefont {Wu}}, \bibinfo {author} {\bibfnamefont {Y.}~\bibnamefont
  {Zhang}}, \bibinfo {author} {\bibfnamefont {A.~C.}\ \bibnamefont {Komarek}},
  \bibinfo {author} {\bibfnamefont {E.}~\bibnamefont {Kampert}},  \emph
  {et~al.},\ }\href@noop {} {\bibfield  {journal} {\bibinfo  {journal}
  {Proceedings of the National Academy of Sciences}\ }\textbf {\bibinfo
  {volume} {115}},\ \bibinfo {pages} {9140} (\bibinfo {year}
  {2018})}\BibitemShut {NoStop}%
\bibitem [{\citenamefont {Cano}\ \emph {et~al.}(2017)\citenamefont {Cano},
  \citenamefont {Bradlyn}, \citenamefont {Wang}, \citenamefont {Hirschberger},
  \citenamefont {Ong},\ and\ \citenamefont {Bernevig}}]{cano2017chiral}%
  \BibitemOpen
  \bibfield  {author} {\bibinfo {author} {\bibfnamefont {J.}~\bibnamefont
  {Cano}}, \bibinfo {author} {\bibfnamefont {B.}~\bibnamefont {Bradlyn}},
  \bibinfo {author} {\bibfnamefont {Z.}~\bibnamefont {Wang}}, \bibinfo {author}
  {\bibfnamefont {M.}~\bibnamefont {Hirschberger}}, \bibinfo {author}
  {\bibfnamefont {N.~P.}\ \bibnamefont {Ong}}, \ and\ \bibinfo {author}
  {\bibfnamefont {B.~A.}\ \bibnamefont {Bernevig}},\ }\href@noop {} {\bibfield
  {journal} {\bibinfo  {journal} {Physical Review B}\ }\textbf {\bibinfo
  {volume} {95}},\ \bibinfo {pages} {161306} (\bibinfo {year}
  {2017})}\BibitemShut {NoStop}%
\end{thebibliography}%

% \cleardoublepage

%%%%%%%%%% Merge with supplemental materials %%%%%%%%%%
\newpage
%\pagebreak
\widetext
\begin{center}
\textbf{\large Supplemental Materials:\\ Signatures of real-space and reciprocal-space topology  in the Eu(Ga$_{1-x}$Al$_x$)$_4$ system}

\end{center}
%%%%%%%%%% Merge with supplemental materials %%%%%%%%%%
%%%%%%%%%% Prefix a "S" to all equations, figures, tables and reset the counter %%%%%%%%%%
\setcounter{equation}{0}
\setcounter{figure}{0}
\setcounter{table}{0}
\setcounter{page}{1}
\makeatletter
\renewcommand{\theequation}{S\arabic{equation}}
\renewcommand{\thefigure}{S\arabic{figure}}
\renewcommand{\thetable}{S\arabic{table}}

\begin{figure}[b]
\includegraphics[width=\columnwidth]{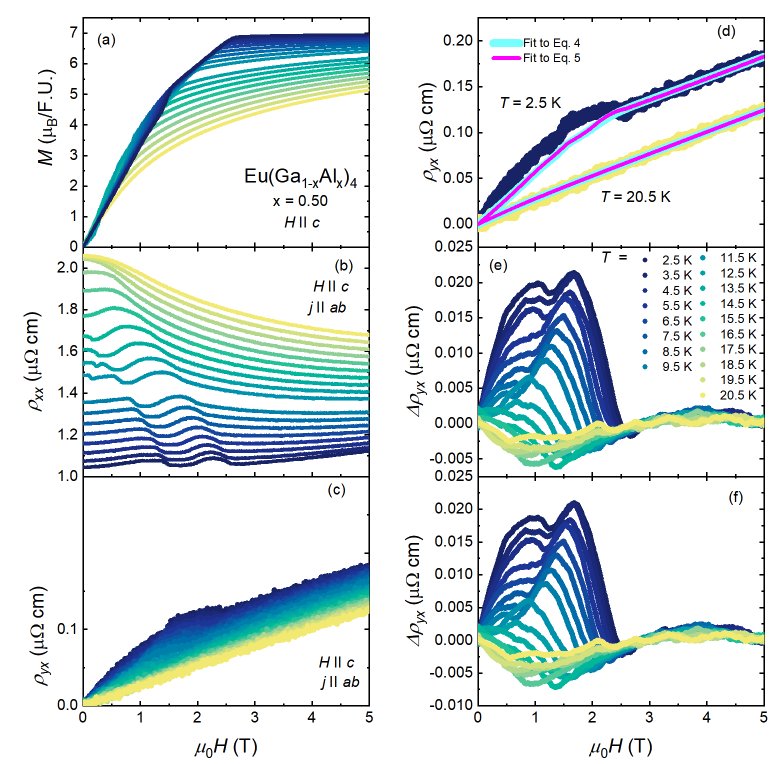}
\caption{\label{THE_5} (a) Isothermal magnetization $M$, (b) longitudinal resistivity $\rho_{xx}$, and (c) Hall resistivity $\rho_{yx}$ measured as a function of magnetic field $\mu_0H$ with field $H~\parallel~c$ at temperatures $2.5~\text{K}~\leq~T~\leq~20.5~$K for \EuGa~x~=~0.50.  Transport measurements are measured with current $j~\parallel~a$. (d) $\rho_{yx}$ at $T~=~2.5~$K (blue symbols) and $T~=~20.5~$K (yellow) with fits to Eq.~\ref{Eq:skewII} (cyan lines) and Eq.~\ref{Eq:intrinsicII} (pink lines). The difference between $\rho_{yx}$ and  fits to (e)  Eq.~\ref{Eq:skewII} or (f) Eq.~\ref{Eq:intrinsicII} resulting in $\Delta\rho_{yx}$ for $2.5~\text{K}~\leq~T~\leq~20.5~$K.}
\end{figure}

\begin{figure}
\includegraphics[width=\columnwidth]{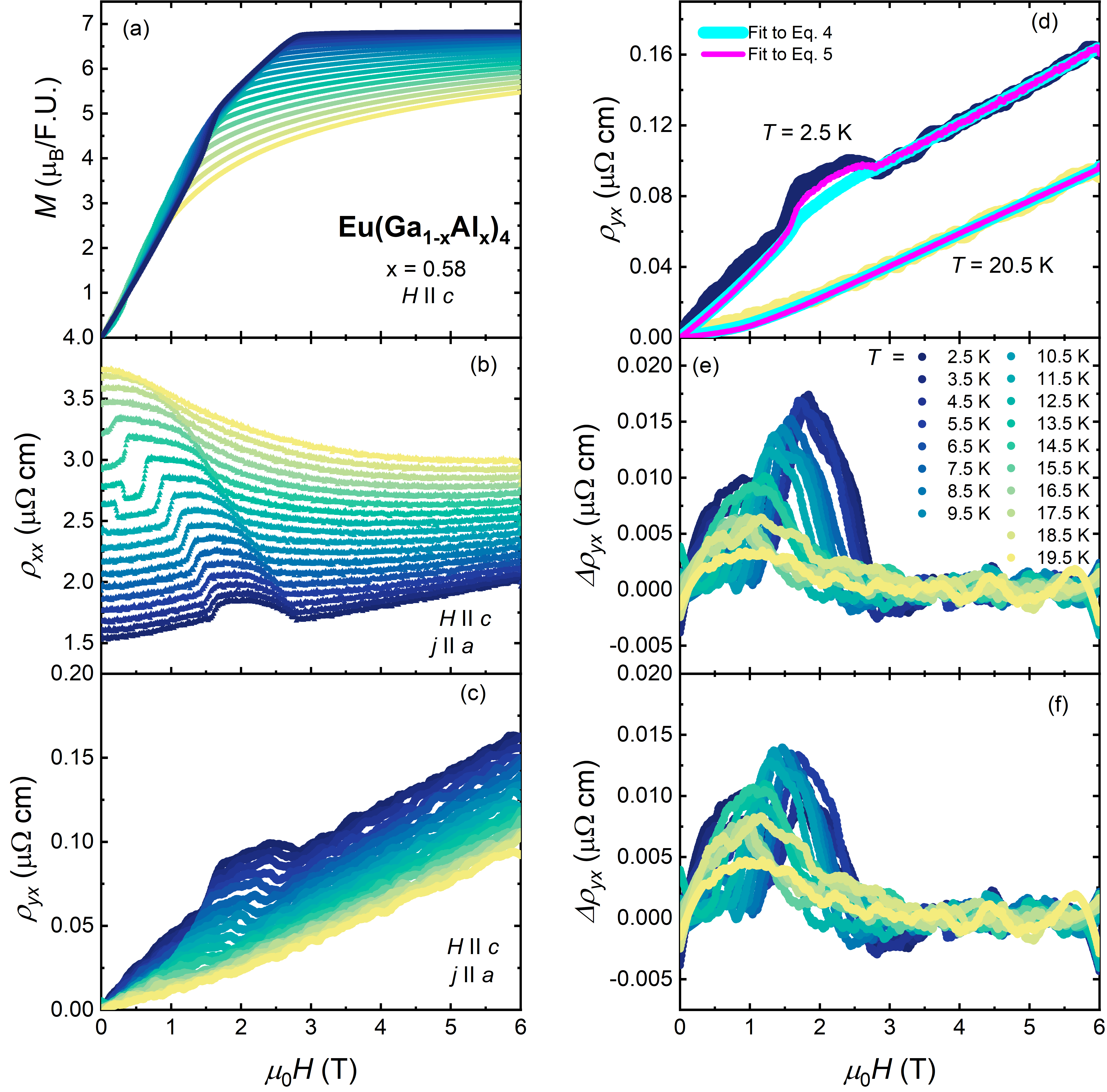}
\caption{\label{THE_58} (a) Isothermal magnetization $M$, (b) longitudinal resistivity $\rho_{xx}$, and (c) Hall resistivity $\rho_{yx}$ measured as a function of magnetic field $\mu_0H$ with field $H~\parallel~c$ at temperatures $2.5~\text{K}~\leq~T~\leq~20.5~$K for \EuGa~x~=~0.58.  Transport measurements are measured with current $j~\parallel~a$. (d) $\rho_{yx}$ at $T~=~2.5~$K (blue symbols) and $T~=~20.5~$K (yellow) with fits to Eq.~\ref{Eq:skewII} (cyan lines) and Eq.~\ref{Eq:intrinsicII} (pink lines). The difference between $\rho_{yx}$ and  fits to (e)  Eq.~\ref{Eq:skewII} or (f) Eq.~\ref{Eq:intrinsicII} resulting in $\Delta\rho_{yx}$ for $2.5~\text{K}~\leq~T~\leq~20.5~$K.}
\end{figure}

\begin{figure*}
\includegraphics[width=\textwidth]{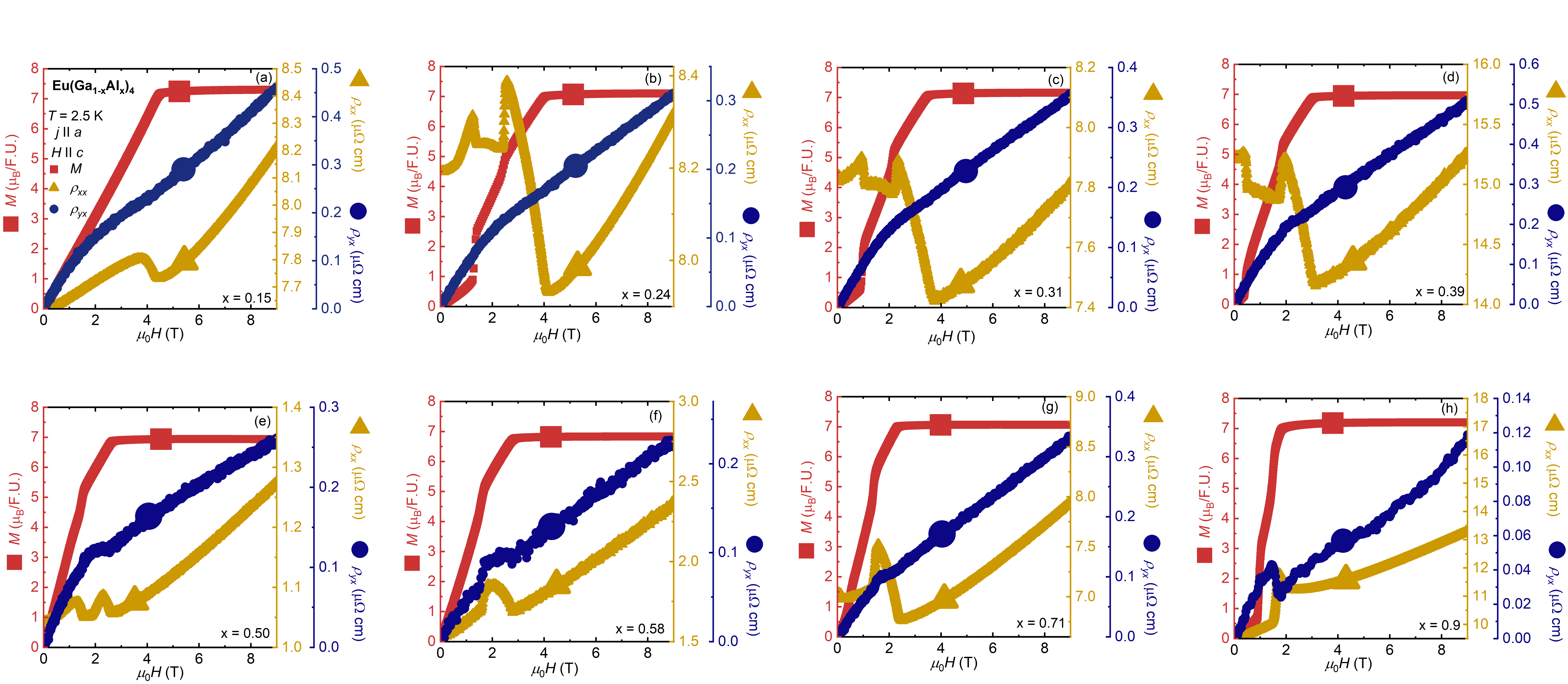}
\caption{\label{two59T} Magnetization $M$ (red squares, left axis), resistivity $\rho_{xx}$ (gold triangles, inner right axis), and Hall resitivity $\rho_{yx}$ (blue circles, outer right axis) as a function of magnetic field $\mu_0H$  measured with magnetic field $H~\parallel~c$ up to 9 T at temperature $T~=~2.5~$K for \EuGa~(a) $x$ = 0.15, (b) $x$ = 0.24, (c) $x$ = 0.31, (d) $x$ = 0.39, (e) $x$ = 0.50, (f) $x$ = 0.58, (g) $x$ = 0.71 and (h) $x$ = 0.90. The transport measurements are measured with the current $j~\parallel~a$.  }
\end{figure*}

\end{document}